\documentclass[aps,prb,amsmath,twocolumn,amssymb,titlepage,superscriptaddress,showpacs]{revtex4}
\usepackage[english]{babel}
\usepackage{graphicx} 
\usepackage{transparent}
\usepackage{amsmath}
\usepackage{amssymb}
\usepackage{epstopdf}
\usepackage{amsfonts}
\usepackage{bm}
\usepackage{times}
\usepackage{pdfpages} 
\usepackage{mathtools}
\usepackage{hyperref}
\usepackage{tabularx}
\usepackage{hhline}

\newcommand{\braket}[2]{\langle #1|#2\rangle}
\newcommand{\matele}[3]{\langle #1|#2|#3\rangle}
\newcommand{\I}{\mathrm{i}}

\newcommand{\ave}[1]{\langle #1 \rangle}

\newcommand{\beq}[1]{\begin{equation} #1 \end{equation}}
\newcommand{\bsplit}[1]{\begin{equation} \begin{split} #1 \end{split} \end{equation}}

\newcommand{\astcycl}{\mathrlap{\kern0.085em{\circlearrowright}}\ast}
\newcommand{\taucycl}{\mathrlap{\kern0.42em{\bullet}}\circlearrowright}

\begin{document}

\title{Multiband nonequilibrium $GW$+EDMFT formalism for correlated insulators}
\author{Denis Gole\v z}
\affiliation{Department of Physics, University of Fribourg, 1700 Fribourg, Switzerland}
\affiliation{Center for Computational Quantum Physics, Flatiron Institute, 162 Fifth Avenue, New York, New York 10010, USA}
\author{Martin Eckstein}
\affiliation{Department of Physics, University of Erlangen-N\"urnberg, 91058 Erlangen, Germany}
\author{Philipp Werner}
\affiliation{Department of Physics, University of Fribourg, 1700 Fribourg, Switzerland}


\begin{abstract}
We study the dynamics of charge-transfer insulators after a photoexcitation using the three-band Emery model which is relevant for the description of 
cuprate superconductors. 
We provide a detailed derivation of the nonequilibrium extension of the multiband $GW$+EDMFT formalism and the corresponding downfolding procedure. 
The Peierls construction of the electron-light coupling is generalized to the multiband case resulting in a gauge invariant combination of the Peierls intraband acceleration and dipolar interband transitions. 
We use the formalism to study momentum-dependent (inverse) photo-emission spectra and optical conductivities. 
The time-resolved spectral function  shows a strong renormalization of the charge-transfer gap and a substantial broadening of some of the bands. While the upper Hubbard bands exhibits a momentum-dependent broadening, an almost rigid band shift is observed for the ligand bands.
The inverse photo-emission spectrum reveals that the inclusion of nonlocal and interband charge fluctuations results in a very fast relaxation of the holes to the top of the Zhang-Rice singlet band. Consistent with the changes in the spectral function, the optical conductivity shows a renormalization of the charge-transfer gap, which is proportional to the photo-doping. The details of the photo-induced changes strongly depend on the dipolar matrix elements, which calls for an \textit{ab-initio} determination of these parameters. 
\end{abstract}

\maketitle
\section{Introduction}
Mott insulators are materials which according to band theory should be metals, but where the motion of the electrons is blocked by the large Coulomb repulsion in relatively localized $d$ or $f$ orbitals. Photo-excitation of charge carriers across the Mott gap, or the dielectric breakdown induced by a strong quasi-static field, results in a nonthermal conducting state. Both types of nonequilibrium phase transitions have been extensively studied experimentally~\cite{iwai2003,perfetti2006,okamoto2007,mayer2015,Yamanaka2017}
and using analytical~\cite{oka2010,zala2012} or computational tools~\cite{eckstein2011,eckstein2013}. While the theoretical works have focused so far mainly on the simple case of a single-band Hubbard model, the experiments are typically performed on materials which would be classified as charge-transfer insulators~\cite{okamoto2007,okamoto2011,cilento2018,peli2017}. In these systems, ligand bands are located in between the upper and lower Hubbard bands, or in the energy range of the lower Hubbard band, and the applied fields may induce holes in the ligand bands, rather than in the lower Hubbard band~\cite{zaanen1985}. For a better understanding of photo-induced nonthermal metallic states in Mott systems and a closer connection between theory and experiment, it is thus essential to extend the previous theoretical studies to charge transfer systems. 

Here, we focus on the dynamics in the $d$-$p$ model for cuprates after field-pulse excitations, which is  a set-up relevant for numerous experimental studies~\cite{okamoto2007,okamoto2011,cilento2018}. Extending the results reported in a recent short study~\cite{golez2019}, we will address several relevant issues which have not been considered in previous single-band model calculations: (i) what is the effect of the pulse energy on the photodoped metal state? Does it make a difference if holes are created in the $p$ bands or in the lower Hubbard band?
 (ii) How do photoinduced changes depend on the excitation protocols, which may include intraband~(Peierls) acceleration and  interband dipolar transitions?
 (iii) To what extent are the band gaps and band dispersions renormalized in the photodoped state? 
  (iv) How are the photoinduced changes in the electronic structure reflected in the optical conductivity?     

In terms of the formalism and numerical effort, the study of the $d$-$p$ model is considerably more involved than a simulation of a single-band Hubbard model with $s$ or $d$ orbitals. In the $d$-$p$ model, we need to treat different bands with different correlation strengths, as well as the dipolar terms appearing in the light-matter coupling. In this work, we will consider a combination of the extended dynamical mean-field theory (EDMFT)~\cite{sun2002} and the $GW$ method\cite{hedin1965} ($GW$+EDMFT).\cite{biermann2003,boehnke2016,ayral2013,huang2014} EDMFT is suitable for describing Mott insulators and allows us to capture strong correlation effects in the $d$ orbitals, while the $GW$ method is appropriate for the description of the more weakly correlated $p$ orbitals, as well as the nonlocal charge fluctuations in the system. In order to assess the relevance of dynamical fluctuations for the description of equilibrium and nonequilibrium states, we furthermore compare the $GW$+EDMFT results to simpler Hartree-Fock+EDMFT (HF+EDMFT) calculations, in which the $p$-orbital interactions are treated at the Hartree-Fock level~\cite{weber2008}.

Apart from its immediate relevance for the theoretical study of photoexcited cuprates, the implementation of a nonequilibrium $GW$+EDMFT scheme for the $d$-$p$ model represents an important step towards the development of {\it ab-initio} simulation approaches for nonequilibrium strongly correlated systems. The formalism captures time-dependent changes in the polarization function and allows to self-consistently compute dynamically screened, effective interaction parameters.\cite{golez2017} This is particularly relevant for the description of excited states of solids, where for example the injection of charge carriers into a Mott insulator completely changes the screening properties.\cite{golez2015} Since $GW$+EDMFT is rooted in the diagrammatic $GW$ formalism, it can furthermore be coupled to $GW$ {\it ab-initio} simulation codes in a way analogous to what has been demonstrated for equilibrium systems in Refs.~\onlinecite{boehnke2016,nilsson2017}. 

Selected results from our $GW$+DMFT investigation of the $d$-$p$ model have been reported in Ref.~\onlinecite{golez2019}.  The purpose of the present work is to provide a detailed description of the formalism, and to present and analyze  polarization- and momentum-dependent data. The paper is organized as follows. In Sec.~\ref{Sec:Model}, we derive the general formalism which combines an EDMFT and $GW$ description of different orbital subspaces, as well as the corresponding downfolding procedure. We then define the $d$-$p$ model and apply the formalism to this specific model. The  coupling of electrons to light for the present multiband problem is derived in Sec.~\ref{Sec:Light}. The equilibrium results are presented in Sec.~\ref{Sec:Equilibrium}, and the dynamics after photoexcitation in Sec.~\ref{Sec:Excitation}. We finish with the conclusions in Sec.~\ref{Sec:Conclusions}.

\section{Model and method}\label{Sec:Model}

\subsection{Multitier $GW$+EDMFT}
The multitier approach\cite{nilsson2017} is motivated by a general property of solid-state systems, where a small subset of orbitals close to the Fermi level are strongly correlated, while higher lying bands are weakly correlated. For example, in the cuprate-inspired models studied in this work, the $p$ orbitals are weakly correlated and can be treated by perturbation theory, while the low-lying $d$ orbitals with strong correlations require a more sophisticated treatment, such as EDMFT. Therefore we separate the full orbital space into strongly and weakly correlated subspaces. To keep the procedure general, we will denote the degrees of freedom in the full space by capital letters, e.g., $A$, those in the correlated subspace by lower case letters, e.g., $a$, and those in higher lying orbitals by lower case letters with an overline, e.g., $\bar a$. The presented formalism is therefore applicable to the general multiorbital case in both subspaces. Only in Sec.~\ref{sSec:dp}, will we restrict the discussion to the specific example of cuprates. 

The formalism applies to a general Hamiltonian of the form
\bsplit{
\label{Eq:Ham}
  H=&H_\text{kin}+H_\text{int}, \\
H_\text{kin}=&\sum_{ij\sigma}\sum_{AB} t_{ij}^{AB} c_{i,A\sigma}^{\dagger} c_{j,B\sigma}-\mu \sum_{iA} n_i^{A},\\
H_\text{int}=&\sum_{i,A} V_{ii}^{AA} n_{i\uparrow}^A n_{i\downarrow}^A+\sum_{i\neq j~\text{or}~A\neq B } V_{ij}^{AB} n_i^A n_j^B,
}
where $i$ and $j$ denote the spatial index of the unit cell and $\mu$ is the chemical potential. We have restricted ourselves to density-density interactions and explicitly separated the local interband interaction term. Here, we have introduced the spin dependent density operators $n_{i\sigma}^{A}=c_{i,A\sigma}^{\dagger} c_{i,A\sigma}$ and the total density operators $n_{i}^{A}=\sum_{\sigma}n_{i\sigma}^{A}$. The on-site energies are fixed by the local-diagonal part of the single-particle term $t_{ii}^{AA}$, and the remaining matrix elements $t_{ij}^{AB}$ define the hopping amplitudes.

The many-body complexity of the system is captured by the Almbladh functional $\Psi[G,W]$ which contains all possible single-particle irreducible diagrams built with the single-particle propagator $G$ and the screened interaction $W$\cite{almbladh1999}.

In general, we aim to construct a conserving approximation for the total system starting from two approximate functionals, one for the correlated subsystem $\Psi_\text{strong}[G,W]$  and one for the higher lying orbitals $\Psi_\text{weak}[G,W]$  as
\begin{align}
  \Psi[G,W]=&\Psi_\text{strong}[G_{ij,ab},W_{ij,ab}]+\Psi_\text{weak}[G_{ij,AB},W_{ij,AB}]\nonumber\\
  &-\Psi_\text{weak}[G_{ij,ab},W_{ij,ab}],
\end{align}
where the last term removes the double counting. In the $GW$+EDMFT approximation, strong correlations are treated within a local EDMFT approximation $\Psi_\text{EDMFT}[G_{ii,ab},W_{ii,ab}]$. The higher-lying orbitals and nonlocal fluctuations are treated within the $GW$ approximation, i.e., the lowest order diagram in the  weak-coupling expansion of $\Psi_\text{$GW$}[G,W]$. The corresponding Almbladh functional is given by\cite{biermann2003,ayral2013}
\begin{align}
 &\Psi_{GW+\text{EDMFT}}[G,W]=\Psi_{\text{EDMFT}}[G_{ii,ab},W_{ii,ab}]\nonumber\\
  &\hspace{5mm}+\Psi_{GW}[G_{ij,AB},W_{ij,AB}]
  -\Psi_{GW}[G_{ii,ab},W_{ii,ab}].
\end{align}
From the stationarity condition we obtain the expression for the self-energy $\Sigma$ and the polarization $\Pi$ as
\beq{
  \Sigma=\frac{\delta \Psi_{GW+\text{EDMFT}}}{\delta G}, \qquad  \Pi=-\frac{\delta \Psi_{GW+\text{EDMFT}}}{\delta W}.
}
Within EDMFT, the local propagators $G_{ii,ab}$ and $W_{ii,ab}$ in the correlated subspace are obtained from the solution of an effective quantum impurity model with action\cite{sun2002,ayral2013,golez2015}
\begin{align}
& S_\text{imp}^{e}[c^*_{a,\sigma},c_{b,\sigma}]=\int_\mathcal{C} dt dt' \nonumber\\
&\Big\{ \sum_{\sigma,ab}c_{a,\sigma}^{*}(t) \left
  [\left((\I \partial_t+\mu ) \delta_{ab} - \delta \mu_{ab}\right) \delta_\mathcal{C}(t,t') -\Delta_{ab,\sigma}(t,t')
  \right ] c_{b,\sigma}(t') \nonumber\\
&- \frac{1}{2}\sum_{ab}  n^a(t)\mathcal{U}_{ab}(t,t') n^b(t') \Big\},
\label{Eq.:Action_EDMFT}
\end{align}
where the retarded interaction $\mathcal{U}$ is coupled to the density operator $n^a$. Here we have separated the self-consistent hybridization function into a time-local part $\delta\mu$, and a contribution $\Delta$ which has no instantaneous term. The origin of the additional shift $\delta \mu$ is that the lattice Hartree-Fock term includes nonlocal contributions, which are not present at the impurity level. The hybridisation function $\Delta(t,t')$ is determined from the
self-consistency condition $G_{ii,ab}=G_{\text{imp},ab}$, and the effective
impurity interaction $\mathcal{U}(t,t')$ is implicitly defined by the condition $W_{ii,ab}=W_{\text{imp},ab}.$
From now on, we will assume a translationally invariant system, although the formalism is easily extended to the spatially inhomogeneous case, and introduce momentum $k$ as a good quantum number. $G$ and $W$ in the full orbital space are given by the fermionic and bosonic Dyson equations 
\beq{
\label{Eq:Dysonlattice_el}
  [G^{-1}_k]_{AB}(t,t')=[(\I\partial_t+\mu)\delta_{AB}-\epsilon_{k,AB}]\delta(t,t')-\Sigma_{k,AB}(t,t'),
  }
\beq{
  [W^{-1}_k]_{AB}(t,t')=[V^{-1}_k]_{AB}\delta(t,t')-\Pi_{k,AB}(t,t'),
  \label{Eq:Dysonlattice_bos}
}
where the dispersion $\epsilon_{k,AB}$ is the Fourier transform of the hopping amplitudes $t_{ij}^{AB}.$
These relations conclude the formal derivation of the theory.

The restriction to density-density interactions considered in this work can be overcome. Here, one should distinguish two cases: i) Additional local interaction terms in the correlated subspace, such as a Kanamori interaction, can be trivially incorporated in a multiorbital extension of the present formalism, as presented in the next section. ii) A treatment beyond density-density interactions outside the local correlated subspace would require a proper EDMFT decoupling and an introduction of new auxiliary fields and two-particle selfconsistency loops. While this is in principle feasible and an important task for the realistic description of correlated materials, it goes beyond the scope of this paper.

\subsection{Downfolding}
While the perturbative $GW$ self-energies and polarizations are explicit functions of $G$ and $W$, $\Sigma^{\text{EDMFT}}$ and $\Pi^{\text{EDMFT}}$ are only implicitly defined through the impurity model. In order to determine the auxiliary quantities $\Delta$ and $\mathcal{U}$, we downfold the electronic and bosonic Dyson equations, see Eqs.~(\ref{Eq:Dysonlattice_el}) and (\ref{Eq:Dysonlattice_bos}), to the correlated subspace, as described in the following paragraph.

\paragraph{Fermionic self-consistency loop.}
We first discuss the downfolding for the electronic Green's function. We can rewrite the fermionic Dyson equation (\ref{Eq:Dysonlattice_el}) by separating the local and nonlocal self-energies
\bsplit{
  &[G_k^{-1}]_{AB}(t,t')=[(\I \partial_t+\mu)\delta_{AB}-(\epsilon_k+\Sigma_{k}^{HF})_{AB}]\delta_\mathcal{C}(t,t')\\
  &-\delta_{Aa}\delta_{Bb}[\Sigma^{\text{EDMFT}}_{ab}(t,t')-\Sigma_{\text{loc},ab}^{GW}(t,t')]-\Sigma_{k,AB}^{GW}(t,t').
}
Here, $\Sigma^{HF}$ is the lattice Hartree-Fock contribution evaluated in the full space.  
The local self-energy of the correlated subspace is described within EDMFT by $\Sigma^{\text{EDMFT}}_{ab}$, and the nonlocal correlations in the entire space are described at the $GW$ level by $\Sigma^{GW}_{k,AB}$.\cite{hedin1965} Since instantaneous contributions can be treated at the level of the effective Hamiltonian, we explicitly separate the Hartree-Fock contribution $\Sigma_{k}^{HF}$ from the higher order EDMFT~($GW$) self-energy contributions $\Sigma^{\text{EDMFT}}$ $(\Sigma^{GW}).$  The term 
$\Sigma_{\text{loc},ab}^{GW}=\frac{1}{N_k}\sum_k\Sigma^{GW}_{k,ab}$ is the local component of the $GW$ self-energy in the correlated subspace, which is subtracted to avoid a double counting of local self-energy contributions in this subspace. Here and in the following we use a short notation for the sum over repeated indices, $\delta_{Aa}X_{ab}\delta_{bB}\equiv \sum_{ab} \delta_{Aa}X_{ab}\delta_{bB}$. In the following, we will separate all  diagonal and local contributions to the self-energy from the rest. This step is a convenient choice for calculations based on strong coupling impurity solvers, where we want to avoid the explicit use of the self-energy in the self-consistency loop, see below for a detailed discussion. We rewrite the lattice Dyson equation as
\begin{align}
 & [G_k^{-1}]_{AB}=[(\I \partial_t +\mu)\delta_{AB}-s_{AB}-
  \Sigma^{\text{EDMFT}}_{ab}\delta_{Aa}\delta_{Bb}]-[\eta_{k}]_{AB}.
  \label{Eq:DysonLattice}
\end{align}
Here, we have separated the local part of the mean-field dispersion, which is now a block-diagonal matrix with strongly and weakly correlated subspaces
\bsplit{
s_{AB}=\begin{pmatrix}
\frac{1}{N_k}\sum_k (\epsilon_{k,ab}+\Sigma_{k,ab}^{HF})  & 0 \\
0 & \frac{1}{N_k}\sum_k (\epsilon_{k,\bar a \bar b}+\Sigma_{k,\bar a\bar b}^{HF}) 
\end{pmatrix}.
}
This separation shifts all the block off-diagonal and block diagonal nonlocal contributions into
\begin{align}
[\eta_k]_{AB}=&[\epsilon_k+\Sigma^{HF}_k]_{AB} - s_{AB}\nonumber\\
& +(\Sigma^{GW}_{k,AB}-\delta_{Bb}\delta_{Aa}\Sigma^{GW}_{\text{loc},ab}). 
\label{Eq:etak}
\end{align}
We note that the local components vanish in the correlated subspace by construction, $\frac{1}{N_k}\sum_k [\eta_k]_{ab}=0.$ 

Once we have solved the lattice Dyson equation, we have to define the auxiliary impurity model (hybridization function $\Delta$ and retarded interaction $\mathcal{U}$) for the correlated subspace. 
The main identity to downfold the Dyson equation from the full space to the correlated subspace is the equation for the inverse of a block matrix
\begin{widetext}
\beq{
  \begin{pmatrix}
  M_{11} & M_{12}\\
  M_{21} & M_{22}
  \end{pmatrix}^{-1}=
  \begin{pmatrix}
  [M_{11}-M_{12}M_{22}^{-1}M_{21}]^{-1} & - [M_{11}-M_{12}M_{22}^{-1}M_{21}]^{-1}M_{12}M_{22}^{-1} \\
  - [M_{22}-M_{21}M_{11}^{-1}M_{12}]^{-1}M_{21}M_{11}^{-1} & [M_{22}-M_{21}M_{11}^{-1}M_{12}]^{-1}
  \end{pmatrix}.\label{Eq:Inverse}
}
\end{widetext}
In order to apply it to the lattice Dyson equation
(\ref{Eq:DysonLattice}) we define an auxiliary propagator for the correlated subspace $g_k$,
\bsplit{
  &[g_k^{-1}]_{ab}(t,t')=\\
  &\quad([\I\partial_t+\mu]\delta_{ab}-s_{ab})\delta(t,t')-\Sigma^{\text{EDMFT}}_{ab}(t,t')
  -\eta_{k,ab}(t,t'),
}
and similarly for the high-lying orbitals
\bsplit{
 \bar g_{k,\bar a\bar b}^{-1}(t,t')=([\I\partial_t +\mu]\delta_{\bar a\bar b}- s_{\bar a\bar b})\delta(t,t')-  \eta_{k,\bar a \bar b}(t,t').
}
Using Eqs.~(\ref{Eq:etak}) and (\ref{Eq:Inverse}), the lattice Dyson equation for the correlated subspace may now be rewritten as
\bsplit{
  &G_{k,ab}^{-1}(t,t')\\
  &=([\I\partial_t+\mu]\delta_{ab}-s_{ab})\delta(t,t')-\Sigma^{\text{EDMFT}}_{ab}(t,t')-\tilde \eta_{k,ab}(t,t'),
\label{Eq:Dyson_downfold}
}
where $\tilde \eta_{k,ab}(t,t')=\eta_{k,ab}(t,t')+\tilde
\Sigma_{k,ab}(t,t'),$ and the auxiliary self-energy is
 \bsplit{ \tilde \Sigma_{k,ab}&=\eta_{k,a\bar b} * g_{k,\bar b \bar c} *
  \eta_{k,\bar cb}.\label{Eq:tildeSigmak} }
Here, $*$ denotes a matrix multiplication in the
orbital sector and a convolution of times. (For HF+DMFT, $\eta_k$ is
instantaneous and the convolutions reduce to multiplications.)  

We determine $\Sigma^{\text{EDMFT}}(t,t')$ by solving the auxiliary impurity problem and imposing the self-consistency in the correlated subspace,
\beq{
\frac{1}{N_k}\sum_k [G_{k}]_{ab}=G_{\text{imp},ab}.  
}
The auxiliary impurity problem fulfills the Dyson equation, which is understood as a matrix equation in orbital space
\bsplit{
  &G^{-1}_{\text{imp},ab}(t,t')\\
  =&\left([\I \partial_t +\mu]\delta_{ab} -\Sigma^{H}_{\text{imp},ab}(t) -\delta\mu_{ab} -\epsilon_{\text{loc},ab} \right) \delta_\mathcal{C}(t,t') \\
  &-\Delta_{ab}(t,t')-\Sigma^{\text{EDMFT}}_{ab}(t,t'),
  \label{Eq:DysonImp}
}
where the impurity Hartree term is given by $\Sigma^{H}_{\text{imp},ab}(t)=\delta_{ab}\int d\bar t~\mathcal{U}_{ac}(t,\bar t) \ave{n^{c}(\bar t)}$ and  the local energy level by $\epsilon_{\text{loc}}=\frac{1}{N_k}\sum_k \epsilon_{k}.$ The additional chemical potential shift $\delta \mu$ corresponds to the difference between the Hartree-Fock contribution on the lattice and on the impurity level,
\begin{align}
  \delta \mu_{ab}(t)&=\Sigma^{HF}_{\text{loc},ab}[G^\text{latt},V]-\Sigma^{H}_{\text{imp},ab}[G^\text{imp},\mathcal{U}]\nonumber \\
  &=s_{ab}(t)-\epsilon_{\text{loc},ab}(t)-\Sigma^{H}_{\text{imp},ab}(t).
  \label{Eq:deltaMU}
\end{align}  

\paragraph{Self-consistency without self-energy.}
Most impurity solvers provide a way to compute the Green’s function rather than the self-energy. In particular, this is the case for the
strong-coupling expansion (NCA) which will be used in this work~\cite{grewe1981,eckstein2010} but also DMRG~\cite{wolf2014,balzer2015} and QMC~\cite{hirsch1986,rubtsov2005,werner2006}. In this case, the determination of the time-dependent self-energy from an inversion of the impurity Dyson equation can be a numerically ill-conditioned problem~\cite{aoki2014_rev}, and it is preferable to close  the self-consistency cycle without explicitly evaluating the self-energy. The main idea is to rewrite the lattice Dyson equation with the auxiliary function $Z$, which is purely local and diagonal and shift the rest into $\eta_k$. Here, we rely on the simple relation that the inverse of a block-diagonal matrix is a block-diagonal matrix, which decouples the evaluation of the auxiliary quantity $Z$ for each subsector. We should stress that any other choice would lead to a coupling between different orbitals and would require an evaluation of the impurity self-energy. To be explicit, we rewrite the lattice Dyson equation as
\begin{align}
  [G_k^{-1}]_{AB}&=
  [(\I \partial_t +\mu)\delta_{AB}-s_{AB}-
  \Sigma^{\text{EDMFT}}_{ab}\delta_{aA}\delta_{bB}]-\eta_{k,AB}\nonumber\\
  &=[Z^{-1}-\eta_{k}]_{AB}.
  \label{Eq:DysonLattice2}
\end{align}
The inverse of the local propagator $Z$ explicitly reads
\bsplit{
\label{eq:Z}  
  Z^{-1}_{AB}=&\delta_{Aa} \delta_{Bb} [(\I\partial_t+\mu)\delta_{ab}-s_{ab}-\Sigma^{\text{EDMFT}}_{ab}] \\
 & + \delta_{A\bar a} \delta_{B\bar b}[(\I\partial_t+\mu)\delta_{\bar a\bar b}-s_{\bar a \bar b}].
}
 The elements $Z_{ab}$ are obtained from the impurity Dyson equation $G^\text{imp}=Z+Z*\Delta*G^\text{imp}$ as a matrix equation in the correlated subspace, while the rest are free propagators. The solution of the lattice Dyson equation is then obtained from  
\beq{
  G_k=Z+Z*\eta_k*G_k,
  \label{Eq:Dyson}
}
which is a matrix equation in the full orbital space. This is the stable form of the Dyson equation, which is used in our numerical implementation. $\eta_k$ is obtained from Eq.~(\ref{Eq:etak}) and in the case of $GW$+EDMFT has an instantaneous and a retarded part. Then, by imposing the self-consistency condition $G_{\text{loc},ab}=G_{\text{imp},ab}$ we obtain the equation for the hybridization function $\Delta$ from Eqs.~(\ref{Eq:Dyson_downfold}) and (\ref{Eq:DysonImp}). In terms of $Z$, these equations read
\bsplit{
  G_{k}^{-1}=&Z^{-1}-\tilde \eta_{k}, \\
  G_{\text{imp}}^{-1}=&Z^{-1}-\Delta+[s-\delta \mu -\epsilon_{\text{loc}} -\Sigma^{H}_{\text{imp}}].
  \label{Eq.:Delta}
}
The term in the square bracket is eliminated by the choice of $\delta \mu$ in Eq.~(\ref{Eq:deltaMU}). 
Then the hybridization function $\Delta$ is determined from Eqs.~(\ref{Eq.:Delta}), in a way analogous to the single-band case:\cite{eckstein2011,aoki2014_rev}
\bsplit{
  G_{k}=&Z+Z * \tilde \eta_{k} * G_{k},\\
  G_{1}=&\frac{1}{N_k}\sum_k \tilde \eta_{k} * G_{k}=\Delta*G_{\text{loc}},\\
  G_{2}=&\frac{1}{N_k}\sum_k [ \tilde \eta_{k} +\tilde \eta_{k} *  G_{k} * \tilde \eta_{k}]= \Delta+\Delta*G_{\text{loc}}*\Delta, 
  \label{Eq.:Delta2}
}
so that $\Delta$ can be calculated by solving 
\begin{equation}
  [1+G_{1}]*\Delta=G_{2}.
   \label{Eq.:Delta3}
\end{equation}
All relations from Eq.~\ref{Eq:Dyson} to Eq.~\ref{Eq.:Delta3} should be understood as block-diagonal matrix equations with blocks corresponding to the strongly correlated and weakly correlated subspaces. The last expression shows that the hybridization function $\Delta$ includes information about direct hopping events to the neighboring sites via $\eta_k$, as well as the hybridization between the two subspaces via $\tilde \eta_{k}.$ Note that $G_{2}$ does not have an instantaneous part, because $\sum_k \tilde \eta_k=\sum_k \tilde \Sigma_k$ has no instantaneous part and in the correlated subspace $\sum_k \eta_k=0$ by construction. Our numerical implementation is based on the stable version of the Dyson equation [Eq.~(\ref{Eq.:Delta3})] and $\tilde \eta_k$ is evaluated using the auxiliary self-energy $\tilde \Sigma_k$ defined in Eq.~(\ref{Eq:tildeSigmak}).

\paragraph{Bosonic self-consistency loop}

We next discuss the downfolding for the screened interaction.  The bosonic lattice Dyson equation is 
\begin{align}
  &[W_k^{-1}]_{AB}=[V_k^{-1}]_{AB}\nonumber\\
  &\hspace{5mm}-[\Pi^\text{EDMFT}]_{ab}\delta_{Aa}\delta_{Bb}-(\Pi_{k,AB}^{GW}-\Pi_{\text{loc},ab}^{GW}\delta_{Aa}\delta_{Bb}),
  \label{Eq:DysonBos}
\end{align}
while the impurity  Dyson equation reads
\beq{
  W_\text{imp}^{-1}=\mathcal{U}^{-1}-\Pi_\text{imp}.
}
We will use these equations and the possibility to calculate the charge susceptibility $\chi_{\text{imp}}$ of the impurity model. The charge susceptibility acts like a $T$-matrix for the bosonic impurity Green's function and allows us to calculate the impurity polarization $\Pi_\text{imp}$ as follows:
$W_{\text{imp}}=\mathcal{U}+W_{\text{imp}}*\Pi_\text{imp}*\mathcal{U}=\mathcal{U}-\mathcal{U}*\chi_{\text{imp}}*\mathcal{U}$, or 
 \begin{align}
 &(1-\chi_{\text{imp}}*\mathcal{U})*\Pi_\text{\text{imp}}=-\chi_{\text{imp}}.
 \label{Eq:Pi}
\end{align}
In the lattice Dyson equation, we set the elements of the lattice polarization $\Pi^{\text{EDMFT}}$ to $\Pi_{\text{imp}}$ within the correlated subspace. The rest of the lattice polarization is filled with the $GW$ polarization, where local components in the correlated subspace have been subtracted due to the double counting. It thus becomes
 \beq{
   [1-V_k (\Pi^{\text{EDMFT}}_{ab}\delta_{Aa}\delta_{Bb}+\Pi_{k,AB}^{GW}-\Pi_{\text{loc},ab}^{GW}\delta_{Aa}\delta_{Bb})]*W_{k}=V_k,
\label{Eq:DysonBosFIN}
}
and from $W_k$ we can calculate the local screened lattice interaction $W_\text{loc}.$ This is once again a stable form of the bosonic Dyson equation, which is employed in our implementation.  The new approximation for the impurity effective interaction $\mathcal{U}$ is obtained by identifying the local screened lattice interaction in the correlated subspace with the screened impurity interaction, $W_\text{loc}=W_{\text{imp}}$:
\beq{
   [1+W_{\text{imp}}*\Pi^{\text{EDMFT}}]*\mathcal{U}=W_{\text{imp}},
   \label{Eq:ImpBoson}
}
which is understood as a matrix equation in the correlated subspace.
\paragraph{Summary of the self-consistency loop}

The full self-consistency loop involves the following steps: 
\begin{enumerate}
\item Start with an initial guess for $\Delta$ and $\mathcal{U}$ and $s_{ab}=\epsilon_{\text{loc},ab} + \Sigma^{HF}_{\text{loc},ab}$.
\item Solve for $G_\text{imp}$ and $\chi_\text{imp}$ in the subspace with the impurity Hartree term $\Sigma^{H}_\text{imp}$ and $\delta \mu= \Sigma^{HF}_{\text{loc}} - \Sigma^{H}_{\text{imp}}$.
\item Invert $Z +Z*\Delta*G_\text{imp} =G_\text{imp}$ to get $Z$ in the correlated subspace, while the $Z_{\bar a\bar b}$ are obtained from the free solution $Z_{\bar a\bar b}=(\I\partial_t+\mu-s_{\bar a\bar b})^{-1}$. The total $Z$ is constructed as in Eq.~(\ref{eq:Z}).
\item Get the local polarization in the correlated subspace by solving Eq.~(\ref{Eq:Pi}).
\item  Compute the lattice propagator $G_k$ ($W_k$) in the full orbital space for all momenta $k$ using the fermionic (bosonic) lattice Dyson equation, see Eq.~(\ref{Eq:Dyson}) (Eq.~(\ref{Eq:DysonBosFIN})). These are both stable versions of the Dyson equation.
\item Optional: Evaluate the nonlocal $GW$ self-energy $\Sigma_{k,AB}^{GW}$ and polarization $\Pi_{k,AB}^{GW}.$
\item Compute the $\bar g_k$ in the weakly correlated subspace
by solving $\bar g_k=\bar g_{k,0}+\bar g_{k,0}*\eta_{k}*\bar g_k$ for $\bar g_k$, where $\bar g_{k,0}^{-1}=\I\partial_t +\mu-s_{\overline a\overline  b}$. As the self-energy in the weakly correlated subspace is explicitly known, this is just an ordinary Dyson equation. Compute the auxiliary self-energy $\tilde \Sigma_k=\tilde \eta_k^{\dagger}\bar g_k \tilde \eta_k$ for the downfolding.
\item Solve Eqs.~(\ref{Eq.:Delta2}) and (\ref{Eq.:Delta3}) 
to obtain the new approximation for $\Delta$ and calculate the updated $\delta \mu$.
\item Obtain the new approximation for the effective impurity interaction $\mathcal{U}$ by solving Eq.~(\ref{Eq:ImpBoson}).
\end{enumerate}

\subsection{\texorpdfstring{$d$-$p$ model}{Lg}}\label{sSec:dp}
We will now apply the formalism introduced in the previous section to the two-dimensional Emery model for cuprates\cite{emery1987}, which includes the correlated Cu $d_{x^2-y^2}$ and higher lying $O$ $p_x$ and $p_y$ orbitals. The position of the $p_x$ $(p_y)$ orbital is shifted with respect to the $d$ orbital by half a unit vector in the $x$ $(y)$ direction. Since the local interactions in the $d$ orbital are stronger than in the $p$ orbitals, we restrict the correlated subspace (EDMFT treatment) to the $d$-orbital, while the $p$ orbitals are treated at the Hartree-Fock or $GW$ level. 

The system Hamiltonian is parametrized by Eq.~(\ref{Eq:Ham}). The on-site energies are given by the local-diagonal component of the single-particle term $t_{ii}^{AA}=\{\epsilon_d,\epsilon_d+\Delta_{pd},\epsilon_d+\Delta_{pd}\}.$ Here, the band separation $\Delta_{pd}$ determines the difference between the on-site energy for the $d$ orbital, $\epsilon_d$,  and the on-site energy for the $p$ orbitals, $\epsilon_d+\Delta_{pd}.$ We denote the nearest neighbor hopping between the $d$ and $p_x$ and $p_y$ orbitals by $t_{ii}^{dp_y}=t_{ii}^{dp_y}\equiv t^{dp}$ and the hopping amplitude between the $p_x$ and $p_y$ orbitals by $t_{ii}^{p_xp_y}\equiv t^{pp}.$ The density-density interaction vertex is given by $V_{ii}^{dd}=U_{dd}$ and $V_{ii}^{dp_x}=V_{ii}^{dp_y}=\frac{1}{2}U_{dp}$ for nearest-neighbor $d$ and $p$ orbitals. 

It is useful to introduce a spinor for the unit cell $i$,
\beq{
  \psi_{i,\sigma} \equiv 
  \begin{pmatrix}
    c_{i, d\sigma} \\
    c_{i, p_x\sigma} \\
    c_{i, p_y\sigma}
\end{pmatrix} ,
}
and its Fourier transform
\beq{
  \psi_{k,\sigma} =\frac{1}{\sqrt{N_k}}\sum_i e^{-\I k R_i} 
  \begin{pmatrix}
    c_{i, d\sigma} \\
    c_{i, p_x\sigma} \\
    c_{i, p_y\sigma}
\end{pmatrix} ,
}
which allows to write the single particle part of the Hamiltonian in the momentum representation as
\begin{widetext}
\label{Eq.:Hkin}
\bsplit{
&H_\text{kin}=\frac{1}{N_k} \sum_{k} \psi_{k,\sigma}^{\dagger} h(k) \psi_{k,\sigma}, \\
&h(k)=\begin{pmatrix}
     2 t_{dd} (\cos(k_x)+\cos(k_y))+\epsilon_d & 2 \I t_{pd} e^{\I k_x/2} \sin(k_x/2) & -2\I t_{pd} e^{\I k_y/2} \sin(k_y/2) \\
    \text{h.c.} & \Delta_{pd}+\epsilon_d & 2 t_{pp}\cos((k_x+k_y)/2) [e^{\I (k_x+k_y)/2}-e^{\I (k_x-k_y)/2}] \\
    \text{h.c.} & \text{h.c.} & \Delta_{pd}+\epsilon_d
\end{pmatrix}.
}  
Here, ``h.c." stands for the Hermitian conjugate part of the matrix. We
note that in some previous studies a Fourier transform with a shifted
position of the $p$ orbitals has been
used.\cite{maekawa2013,hansmann2014}

By performing the same Fourier
transform as for the single particle part of the Hamiltonian we arrive at the following momentum representation of the interaction
\bsplit{
V^{AB}_k=\begin{pmatrix}
  U_{dd}/2 &   U_{dp} e^{-\I k_x/2} \cos(k_x/2)  &  U_{dp}e^{-\I k_y/2}\cos(k_y/2)  \\
   U_{dp} e^{\I k_x/2}\cos(k_x/2)  & U_{pp}/2  & 0       \\
   U_{dp} e^{\I k_x/2}\cos(k_y/2)  & 0       & U_{pp}/2  \\
\end{pmatrix},
}
\end{widetext}
where special care needs to be taken for the local and orbital diagonal components to fulfill the Pauli exclusion principle, see Sec.~\ref{Sec.:Diagrammatic}. We solve this lattice problem using the $GW$+EDMFT method \cite{biermann2003,ayral2013,huang2014,golez2017} introduced above, which is based on EDMFT.\cite{sun2002,ayral2013,huang2014,golez2015} The downfolding is thus performed from the  three-orbital space of the full model to the $d$ orbital correlated subspace both for the electronic Green's function and the (bosonic) screened interaction.   

A nontrivial problem in practice is the convergence. Because the Hartree and Fock terms are quite large, the solution can jump from fully occupied to fully unoccupied $d$ orbitals during the self-consistency cycle. This is why we fix the total occupation to $n$ and the relative occupation between the orbitals $r=n_p/n_d,$ which determines the local-diagonal part of the density matrix and the local Hartree and Fock shifts. 
$r$ is determined by minimizing the difference between the fixed and calculated local density matrix at a given chemical potential. Once a sufficiently good approximate solution is obtained, we release the constraint on the local block-diagonal matrix.

The electronic propagator is defined as 
\begin{align}
 G_{k,\sigma}(t,t')=& -\I \ave{\mathcal{T}_\mathcal{C} \psi_{k\sigma}(t) \psi_{k\sigma}^{\dagger}(t')} \nonumber\\
=& \begin{pmatrix}
  G_{k,dd,\sigma}    & G_{k,dp_x,\sigma}   & G_{k,dp_y,\sigma}   \\
  G_{k,p_xd,\sigma}  & G_{k,p_xp_x,\sigma} & G_{k,p_xp_y,\sigma} \\
  G_{k,p_yd,\sigma}  & G_{k,p_yp_x,\sigma} & G_{k,p_yp_y,\sigma},
  \end{pmatrix},
\end{align}
with $\langle \ldots \rangle = \frac{1}{Z}\text{Tr} [\mathcal{T}_\mathcal{C}e^{\I S_\text{latt}}\ldots]$ and $S_\text{latt}$ the lattice action, $Z$ the partition function of the initial equilibrium state with inverse temperature $\beta$, and $\mathcal{T}_\mathcal{C}$ the time ordering operator on the Kadanoff-Baym contour which runs from time $0$ to time $t$ along the real time axis, back to time zero and then to $-\I \beta$ along the imaginary-time axis.\cite{aoki2014_rev} The free propagator is similarly obtained from the single-particle Hamiltonian, or the corresponding lattice action. For the matrix elements of the tight-binding approximation, see Ref.~\onlinecite{hansmann2014}.  
Since there is hopping between all orbitals in the unit cell, the free propagator is a full matrix. 

The bosonic propagator is defined as 
\begin{align}
 W_{k,\sigma}(t,t')=&-\I \ave{\mathcal{T}_\mathcal{C} \phi_{k\sigma}(t) \phi_{k\sigma}^{\dagger}(t')} \nonumber\\
=& \begin{pmatrix}
  W_{k,dd,\sigma}    & W_{k,dp_x,\sigma}   & W_{k,dp_y,\sigma}   \\
  W_{k,p_xd,\sigma}  & W_{k,p_xp_x,\sigma} & W_{k,p_xp_y,\sigma} \\
  W_{k,p_yd,\sigma}  & W_{k,p_yp_x,\sigma} & W_{k,p_yp_y,\sigma}
\end{pmatrix},
\end{align}
where the spinor  $\phi_{k\sigma}=\{\phi_{k,d\sigma}$, $\phi_{k,p_x\sigma}$, $\phi_{k,p_y\sigma}\}$ originates from a Hubbard-Stratonovich decoupling of all the interaction terms in $H_\text{int}$.

\subsection{Self-energy and Polarization diagrams}\label{Sec.:Diagrammatic}
In a conserving approximation, the selfenergy and polarization are
given by the derivative of the Luttinger-Ward functional with respect to the Green function and screened interaction, respectively. 
We will provide here the expressions for the mean-field (nonretarded Hartree+Fock) and the retarded $GW$ approximations. 

\paragraph{Hartree}
The  Hartree expression for the Emery model is given by
\bsplit{
  \Sigma^H_{\alpha\alpha}(t,t')=&
  \delta(t,t')
  \sum_{\beta\neq\alpha}\rho_{\beta\beta}V^{\alpha\beta}_{k=0},
\label{Eq:Hartree}
}
where we have introduced the single-particle density matrix
$\rho_{k,\alpha\beta}=\ave{c^{\dagger}_{\beta,k} c_{\alpha,k}}$, its local value $\rho_{\alpha\beta}=\frac{1}{N_k}\sum_{k}\rho_{k,\alpha\beta}$, and the combined index $\alpha=(A,\sigma),$ where $A$ is the band and $\sigma$ the spin index. As marked in the sum we need to fulfill the Pauli exclusion principle and therefore we only sum over distinct indices $\beta\ne \alpha$.
For the $d$ orbital the expression is given by
\bsplit{
  &\Sigma^H_{dd,\sigma}(t,t')=\\
  &[V^{dd}_\text{loc}  \rho_{dd,\bar \sigma}+V^{dd,\text{nl}}_{k=0}  \rho_{dd}+V^{dp}_{k=0} \rho_{p_xp_x}+V^{dp}_{k=0} \rho_{p_yp_y}]\delta(t,t'),
}
where $V^{dd,\text{nl}}$ marks a possible nonlocal $d$-$d$ interaction. We have assumed spin symmetry and used the notation $\rho_{AB}=\sum_{\sigma} \rho_{AB,\sigma}$. 

Now, we can calculate the time-dependent shift of the chemical potential for the impurity problem (see Eq.~(\ref{Eq:deltaMU})), where the mean-field contributions are given by
\begin{align}
\Sigma^{HF}_{\text{loc},dd,\sigma}&=
V^{dd}_\text{loc} \rho_{dd,\bar \sigma}(t) +V^{dd,\text{nl}}_{q=0} \rho_{dd}(t) \nonumber\\ 
&\quad+4V^{dp_x}_{q=0}\rho_{p_xp_x}(t)+4V^{dp_y}_{q=0}\rho_{p_yp_y}(t) , \\
\Sigma^{H}_{\text{imp},dd,\sigma}&= U_{dd} \rho_{dd,\bar \sigma}(t)  + \textstyle{\int} d\overline t \mathcal{D}(t,\overline t)\rho_{dd}(\bar t).
\end{align}
Here we have introduced the retarded part of the effective impurity
interaction as $\mathcal{D}=\mathcal{U}-V^{dd}_\text{loc}$. We can get rid of the term $\textstyle{\int} d\overline t\mathcal{D}(t,\overline t)\rho_{dd}(\overline t)$ by performing an  expansion in the
density fluctuations 
$\bar n=n-\langle n\rangle$
instead of $n$, see discussion in Ref.~\onlinecite{golez2015}.

\paragraph{Fock}
In order to fulfill the Pauli exclusion principle we should remove the
local diagonal part of the Fock diagram,  
\bsplit{
  &\Sigma^F_{k,\alpha\beta}(t,t)=\\
  &-\frac{1}{N_k}\left[ \sum_q \rho_{k-q,\beta\alpha}(t)  V^{\alpha\beta}_q - 
    \delta_{\alpha\beta}\rho_{\text{loc},\alpha\alpha}(t)  V^{\alpha\alpha}_\text{loc} \right],
}
where $V^{\alpha\alpha}_\text{loc}$ is the local intraorbital interaction and repeated indices are understood as a piecewise product. To give a concrete example, for the diagonal $d$ component this becomes
\beq{
  \Sigma^F_{k,dd,\sigma}(t)=-\frac{1}{N_k}\sum_q V_{q}^\text{dd,nl}(t) \rho_{k-q,dd,\sigma}(t,t^+),
}
where we included the possibility of a nonlocal $d$-$d$ interaction. The off-diagonal parts of the self-energy are not modified, e.~g.
\beq{
  \Sigma^F_{k,dp_x,\sigma}(t)=-\frac{1}{N_k}\sum_q V_{q}^{dp}(t) \rho_{k-q,dp_x,\sigma}(t).
}

\paragraph{$GW$}\label{sSec:GW}
The $GW$ self-energy is given by
\bsplit{
  \Sigma^{GW}_{k,\alpha\beta}(t,t')=
  \frac{1}{N_k}\sum_q \I G_{k-q,\beta\alpha}(t,t')  W_{q,\alpha\beta}(t,t'),
}
from which we have to subtract the local term already present in the impurity self energy. The explicit expression for the diagonal $d$ contribution is
\bsplit{
  \Sigma^{GW}_{k,dd,\sigma}(t,t')=\frac{\I}{N_k}\sum_q G_{k-q,dd,\sigma}(t,t')  W_{q,dd}(t,t'),
}
while the off-diagonal contributions are given by
\beq{
  [\Sigma^{GW}]_{q,dp_x,\sigma}(t,t')=\frac{\I}{N_k}
  \sum_q G_{k-q,dp_x,\sigma}(t,t') W^{q,dp_x}(t,t').
}

The polarization is calculated in an analogous fashion,   
\begin{align}
  \Pi^{GW}_{k,\alpha\beta}(t,t')=\frac{-\I}{N_k}\sum_{q\sigma} G_{k-q,\alpha\beta}(t,t') G_{q,\beta\alpha}(t',t),
\end{align}
where the sum is done only through spin indices $\sigma$ for combined indices $\alpha$ and $\beta$.
The $d$-$d$ and $d$-$p$ components thus are
\begin{align}
  \Pi^{GW}_{k,dd}(t,t')= \frac{-2\I}{N_k} \sum_{q\sigma} G_{k-q,dd,\sigma}(t,t') G_{q,dd,\sigma}(t',t)
\end{align}
and 
\begin{align}
  &\Pi^{GW}_{k,dp_x}(t,t')= \frac{-2\I}{N_k}
  \sum_{q\sigma} G_{k-q,dp_x,\sigma}(t,t') G_{q,p_xd,\sigma}(t',t),
\end{align}
and analogous for all the other terms. From this it follows that the polarization matrix is a full matrix.

\section{Coupling with light}\label{Sec:Light}
In order to simulate the photo excitation, we need to
understand how light couples to matter in the multiband
case. 
In principle, one can obtain a light-matter Hamiltonian by projecting the continuum theory onto a given set of bands. There is however a well-known gauge ambiguity:\cite{loudon2000} While the full theory can be described equivalently in different gauges for the electromagnetic fields, the description resulting from a projection to a given set of bands or Wannier orbitals can depend on the gauge in which the projection is performed. In particular, different descriptions are obtained by projecting the continuum Hamiltonian in the Coulomb gauge, which corresponds to the replacement $\vec p\rightarrow\vec p -e\vec A $, or in the dipolar gauge, where the coupling is between a polarization density and the electric field. This ambiguity is actively discussed for atomic Hamiltonians in the context of cavity quantum electrodynamics.\cite{Bernardis2018b,DiStefano2019} 
\\
In the present case, we are interested in the semiclassical description (the fields are treated classically). Nevertheless, the question of the optimal light-matter Hamiltonian becomes relevant for strongly driven systems, beyond linear response. A common strategy is to request the multiband light-matter Hamiltonian to have a gauge structure even after projection to a given manifold of bands.\cite{boykin2001} This leads to a Hamiltonian which in general includes both multipolar magnetic and electric matrix elements, and Peierls phases depending on the vector potential. Here we give a compact summary of the corresponding construction, which from the outset takes into account the main simplifications which are appropriate for the description of optically driven solids: In particular, all fields are assumed to be spatially homogeneous throughout the solid ($q=0$ approximation). This is suitable in the long wavelength limit; it effectively neglects magnetic dipolar matrix elements,  which are relatively small  even for electric fields of the order of MV/m.
\\
Before presenting this construction, we remark that the $q=0$ approximation can be dropped relatively easily, and one can instead obtain the same light matter Hamiltonian by only neglecting the variation of the fields on the {\em atomic} scale, derive a Hamiltonian with spatially varying fields, and restrict the fields to $q=0$ in the end. Moreover, by making the semiclassical approximation from the outset one does not capture induced long-range dipolar interactions and Lamb-shifts. However, in the present case all interactions (including the long-range Coulomb interaction) are approximated by an effective short-range interaction, whose ab-initio determination is challenging even in the case without light-matter coupling. An alternative but more lengthy derivation of the Hamiltonian, which leads to the same Hamiltonian as below, without making the $q=0$ and semiclassical approximation from the outset, can be found in Ref.~\onlinecite{jiajun2019}.
\\
For the construction of the light-matter Hamiltonian, we start with the semiclassical coupling (which is obtained from the minimal coupling Hamiltonian using a Power-Zienau-Wolley~(PZW) transformation, see Ref.~\onlinecite{loudon2000})
\beq{ \label{gegefeg}H_{ED}=e\sum_{j} \vec r_j \cdot \vec
  E=\vec E \cdot \mathcal{\vec P}, } where we have introduced the dipolar
moment $\mathcal{\vec P}= e\sum_j \vec r_j$, $\vec E$ is the (classical) and spatially homogeneous transversal component of the electric field, $\vec r_j$ is the position of the $j$-th electron and $e$ the electric charge. We then make the  following approximations:
\begin{enumerate}
\item Project the continuum Hamiltonian to a set of localized Wannier basis functions and explicitly separate intra- and inter-unit-cell contributions.
\item As the inter-unit-cell terms break the translational invariance, we apply a unitary transformation which restores the translational symmetry at the expense of time-dependent phase factors. This step is a generalization of the Peierls substitution. The transformation of the inter-unit cell terms will reintroduce the vector potential $\vec A$ into the description. The mixed representation with the electric field $\vec E$ and the vector potential $\vec A$ is the main result of this section. It allows the modeling of translationally invariant systems with intra-band acceleration and inter-orbital dipolar excitations.
\end{enumerate}
In more detail, the projection of \eqref{gegefeg} to a set of Wannier orbitals is obtained as follows: The unit cell may contain several atoms
described by Wannier functions, which we denote by
$\braket{\vec r}{k,n}=\phi_{n}(\vec r -\vec R_{k})$ for the $k$-th unit cell and $n$-th orbital in the unit cell. We thus obtain the matrix elements of the position operator as:
\begin{widetext}
\begin{align}
  \matele{in}{\vec r}{km}&=\matele{in}{\vec r-\vec R_{k}}{km}+\matele{in}{\vec R_{k}}{km}=\vec D_{ik}^{nm}/e+\matele{in}{\vec R_{k}}{km}=\vec D_{ik}^{nm}/e+\vec R_{k} \delta_{k,i} \delta_{n,m}
  \label{Eq:Dipol}
\end{align}
\end{widetext}
where the vector $\vec R_{k}$ represents the center of mass in the unit cell $k$. We have introduced the dipolar matrix element as $\vec D_{ik}^{nm}=e \matele{in}{\vec r-\vec R_{k}}{km}.$

The tight-binding Hamiltonian on the lattice can now be written as
\beq{
  H_{ED}=\sum_{\substack{ij\\nm}} t_{ij}^{nm} c_{in}^{\dagger} c_{jm}+ e\vec E \sum_{\substack{i}} n_i \vec R_{i} +
   \vec E\sum_{\substack{ij\\nm}}  \vec D_{ij}^{nm} c_{in}^{\dagger} c_{jm}.
}
The inter-unit-cell contributions to the Hamiltonian  $e\vec E \sum_{\substack{i}} n_i \vec R_{i}$ break the translational invariance. In order to restore the translational symmetry, we perform the unitary transformation  $U=e^{\I e\sum_{j} \chi_{j} n_j}.$ This transformation removes the  $e \vec E \sum_{j} n_j \vec R_{j}$ term and introduces time-dependent phase factors as in the Peierls substitution:
\begin{align}
  \tilde H_{ED}&=U^{\dagger} H_{ED} U -\I U^\dagger \dot U \nonumber \\
  &=\sum_{ij} t_{ij}^{nm} e^{-\I e(\chi_{i}-\chi_{j})} c_{in}^{\dagger} c_{jm}+e\vec E \sum_{i}n_i \vec R_i \nonumber \\
  &+e\sum_{i} \dot \chi_{i} n_i
  +\vec E \sum_{ij} \vec D_{ij}^{nm} 
  e^{-\I e (\chi_{i}-\chi_{j})} c_{in}^{\dagger} c_{jm},
\end{align}
from where we get the condition $\dot \chi =-\vec E(t) \cdot \vec R_i \rightarrow \vec A(t)=-\int^t \vec E(s) ds$ in order to cancel the spatially inhomogeneous electric field contributions. The Peierls phases are thus given by $\chi_{ij}=\chi_{i}-\chi_{j}=\int_{\vec R_j}^{\vec R_i} \vec A(x) d\vec x=(\vec R_i -\vec R_j) \vec A$. At this point, the integral is independent of the path, as the vector potential is homogeneous and can be written as a gradient field.  The final Hamiltonian is now given by
\beq{
  \tilde H_{ED}=\sum_{ij}t_{ij}^{nm} e^{-\I e\chi_{ij}} c_{in}^{\dagger} c_{jm} +
  \vec E\sum_{ij} \vec D_{ij}^{nm} e^{-\I e\chi_{ij} } c_{in}^{\dagger} c_{jm}.
  \label{Eq:HED}
}
The Peierls transformation yields time-dependent hoppings, but also the nonlocal dipolar matrix elements have a time-dependent phase. Up to this point, the derivation was rather general (up to neglecting quantum effects and making the $q=0$ approximation). This Hamiltonian can now be restricted to a given subset of bands. The important point is that the Hamiltonian still has an exact gauge structure, i.e., the description is invariant under the transformation $\vec{A}\to\vec{A}+\vec{\nabla}\Lambda(\vec{r},t)$, $c_j\to c_j e^{i \Lambda(\vec{R}_j,t)}$, $\phi(\vec{r})\to \phi(\vec{r})-\partial_t\Lambda(\vec{r},t)$
  when a scalar potential term $H_\phi=e\sum_{j,n} \phi(\vec{R}_j)c_{jn}^{\dagger} c_{jn}$ is added, which is the requirement stated in Ref.~\onlinecite{boykin2001}. 
 The real-world application however deserves several comments.
\begin{enumerate}
\item Some ambiguity in this truncation comes from the definition of the dipolar moment, which depends on the choice of the origin $\vec R_{k}$ in Eq.~(\ref{Eq:Dipol}). This issue will be important for an {\it ab-initio} determination of the dipolar matrix elements, see the discussion below, which goes beyond the scope of the current paper. 
\item In the current work, we will truncate the range of the dipolar matrix element to the nearest neighbor. However, this approximation is not crucial and can be relaxed.
\item 
It is a nontrivial question whether the dipolar light-matter Hamiltonian is optimal for a quantitative description of strongly driven phenomena beyond linear response, or whether, e.g., working with a projected continuum Hamiltonian in  a fixed gauge can give quantitatively better results for a given number of bands. A general answer to this question is beyond the scope of this paper. An explicit comparison for sub-cycle driven dynamics in a single band shows that the dipolar Hamiltonian is not only favorable because of its gauge structure, but it also converges more rapidly with the number of bands taken into account.\cite{jiajun2019}
A future combination with {\it ab-initio} input  and its comparison with experimental data could serve as a-posteriori justification. 
\end{enumerate}
As this is the first work which addresses the nonlinear response in correlated systems with both intraband and interband terms, we will focus on their qualitative contributions to physical observables.

Given $\tilde H_{ED}$, the corresponding currents can be obtained as $ \vec j=\partial_t \mathcal{\vec {\tilde P}}(t)=-\I
[\mathcal{\vec{\tilde P}},\tilde H_{ED}]=-\I
[\mathcal{\vec{\tilde P}}, H_t+H_D],$ where we have separated the pure Peierls term $H_t$ and the dipolar term $H_D.$ The polarization is given by
 $\mathcal{\vec{\tilde P}}=\mathcal{\vec P_{\text{dia}}}+\mathcal{\vec P_{\text{off}}}=e \sum_{i} \vec R_i n_i + \sum_{ij\\nm} \vec D_{ij}^{nm} c_{in}^{\dagger} c_{jm} e^{-e\I \chi_{ij}}$.  
The commutator of the diagonal term $\mathcal{\vec P_{\text{dia}}}$ with the kinetic part of the interaction gives the usual contribution from the electron hopping
\begin{align}
 \vec j_K=&-\I [\mathcal{\vec P_{\text{dia}}}, H_t]=\nonumber\\
 &\frac{ -\I e}{N_k} \sum_{\substack{ijl\\ nmp}} t_{ij}^{nm} e^{-e\I\chi_{ij}} \vec R_{l}
  [c_{lp}^{\dagger} c_{lp},c_{in}^{\dagger} c_{jm}]+\text{h.c.}\nonumber\\
&=\I e \sum_{\substack{ij\\nm}} 
(\vec R_{i}-\vec R_{j}) t_{i j}^{nm} e^{-e\I
\chi_{ij}} c_{in}^{\dagger}
c_{jm} + \text{h.c.},
\end{align}
or in momentum space
\begin{align}
\label{Eq:current_Peierls}
&\I e \sum_{\substack{ij\\nm}} t_{ij}^{nm} e^{-\I e\chi_{ij}} (\vec R_{i}-\vec R_{j}) c_{in}^{\dagger} c_{jm} + \text{h.c.}  =\nonumber\\
&\hspace{14mm}
\frac{e}{N_k}\sum_{\substack{k\\nm}} \frac{\partial}{\partial_{\vec k}} \epsilon_{\vec k-e\vec A}^{nm}  c_{kn}^{\dagger}
c_{km}.
\end{align}
The commutator of the diagonal term $\mathcal{\vec P_{\text{dia}}}$ with the dipolar part of the Hamiltonian evalutes to 
\bsplit{
\vec j_{D1}&=-\I [\mathcal{\vec P_{\text{dia}}},H_{D}] \\
&=  \frac{-\I e}{N_k} \sum_{\substack{ijl\\nmp}}  (\vec
E \cdot \vec D)^{nm}_{ij} e^{-\I e\chi_{ij}} \vec R_{l} [c_{lp}^{\dagger} c_{lp}, c_{in}^{\dagger} c_{jm}]+\text{h.c.}\\
&=\I e  \sum_{\substack{ij\\nm}} (\vec E \cdot \vec D)^{nm}_{ij} e^{-\I e\chi_{ij}}  (\vec R_{i}-\vec
R_{j}) c_{in}^{\dagger} c_{jm} +\text{h.c.} 
\label{Eq:current_dipol}
}
and expresses the change in the polarization due to the dipolar transition.
In momentum space, this expression becomes
\bsplit{
-&\I e\sum_{\substack{ij\\nm}} (\vec E \vec D)^{nm}_{ij} e^{-\I e \chi_{ij}}  (\vec R_{i}-\vec R_{j}) c_{in}^{\dagger} c_{jm} +\text{h.c.}  \\
& =-\frac{\I e}{N_k}\sum_k\frac{\partial}{\partial\vec k}
 \Big[\sum_{nm} (\vec E \cdot \vec D)^{nm}_{ij} e^{-\I e (\vec k-e\vec A) (\vec R_{i}-\vec R_{j})}\Big] c_{kn}^{\dagger} c_{km}.
}

The commutator of the off-diagonal polarization $\mathcal{\vec P_{\text{off}}}$ with the kinetic term $H_t$ reduces to 
\beq{
  \vec j_{D2}=\I[\mathcal{\vec P_{\text{off}}},H_t]=-\I e \sum_{\substack{ijk\\nmp}} \vec D_{ki}^{pn}  t_{ij}^{nm} e^{-\I e (\chi_{ij}+\chi_{ki})}  c_{kp}^{\dagger} c_{jm} + \text{h.c.} .
}
If the dipolar matrix element is local in space this expression reduces to  
\beq{
  \vec j_{D2}=-\I e \sum_{\substack{ij\\nmp}} \vec D_{ii}^{pn}  t_{ij}^{nm} e^{-\I e \chi_{ij}} c_{ip}^{\dagger} c_{jm} + \text{h.c.}
}
or in momentum space 
\beq{
  \vec j_{D2}=-\I e \sum_{k nmp}  [\sum_{ij}  \vec D_{ii}^{pn} t_{ij}^{nm} e^{\I (\vec k-e\vec A) (\vec R_{i}-\vec R_{j})}]  c_{kp}^{\dagger} c_{km} + \text{h.c.} .
}

The Peierls $\vec j_K$ and the dipolar $\vec j_{D2}$ contributions to the current are independent of the electric field which makes them dominant at the linear-response level. The dipolar contribution to the current $\vec j_{D1}$ contains a factor of $\vec E$ and therefore it will only contribute at the nonlinear level. In the practical calculation, we evaluate the Peierls contribution to the current  $\vec j_{K}$ by Eq.~(\ref{Eq:current_Peierls}) and the total contribution to the dipolar current by a numerical derivative of the inter-band polarization $\vec j_{D1}+\vec j_{D2}=\partial \mathcal{\vec P_{\text{off}}}/\partial t.$

The dipolar matrix element is a material-specific quantity. In the case of the $d$-$p$ model we will treat only the nearest-neighbor $d$ to $p$ dipolar matrix elements and neglect the rest: \bsplit{ D \equiv&
  \vec D_{ii}^{p_xd}=
  -\vec D_{ii}^{p_yd} \\
  &=e\int d\vec{r}
  \phi_{p_x}(\vec r-R^*) \vec r \phi_{d}(\vec r-R^*),}
  where $R^*$ is the center of mass in the unit cell. For the dipolar matrix element between the $d$ and $p_x$ orbital the time-dependent phase
is $e^{-\I e A_x a_x/2},$ since the distance between the $d$ and $p_x$
orbital is $a/2.$ The same expression holds for the $d$-$p_y$ element, but due to the
phase dependence of the orbital it has the opposite sign. 

\begin{figure}[t]
\includegraphics{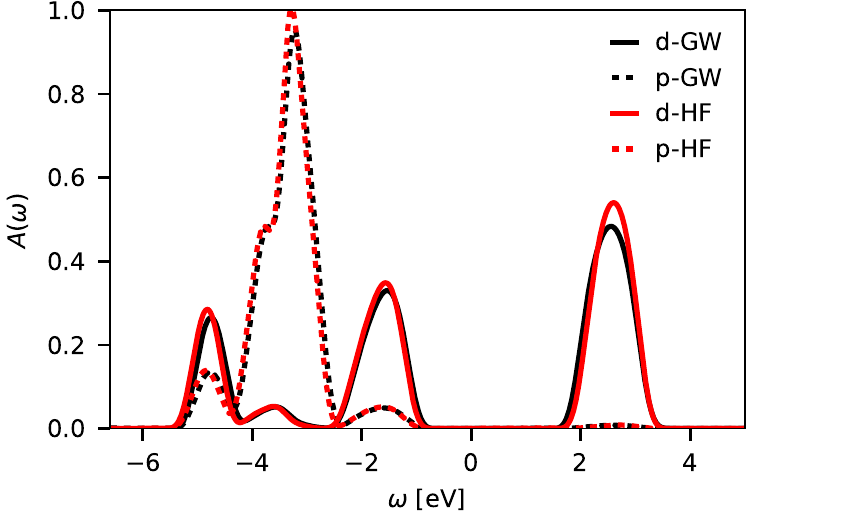}\\
\caption{Orbital-resolved spectral function 
  obtained in the $GW$+EDMFT (black) and HF+EDMFT (red) approximation. The full (dashed) lines represent the $d$ ($p_x$ and $p_y$) orbitals.}.
\label{Fig:Spectral}
\end{figure}

To the best of our knowledge, the dipolar matrix element between the
$d$ and $p$ orbital has not been reported in the literature. For the purpose of this work, we have performed the integral numerically using a cluster of $d_{x^2-y^2}$ and $p_{x},p_{y}$ orbitals, whose spatial
profile is given by a Gaussian function with a variable width
$\sigma$. The evaluation of the dipolar matrix element for a Gaussian
with width $\sigma=0.01 a,$ where $a$ is the lattice
constant, yields a dipolar matrix element
$|D|\approx 0.3$ $e$a, with $e$ the electronic charge. The matrix element strongly depends on the spatial
width of the orbital and in the following we will use two
characteristic values, namely $|D|=0.1,0.3$ $e$a. For a quantitative
description of the photoexcitation in multiband materials, it will be
crucial to determine these parameters from {\it ab-initio} calculations, 
taking into account material-specific properties.

\begin{figure}[t]
\includegraphics{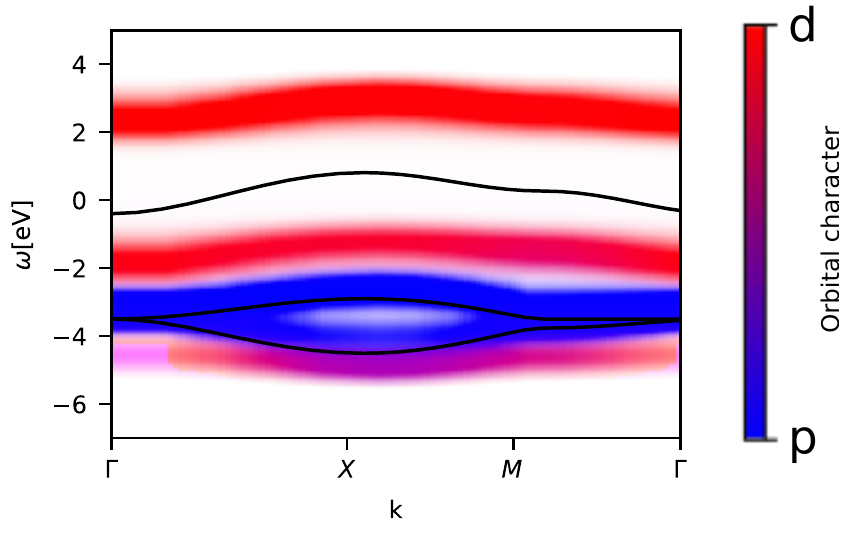}
\caption{Momentum-resolved spectral function $A_{k}(\omega)$. The color coding indicates the orbital contribution to the spectral weight (red for $d$- and blue for $p$-orbital character). The black lines correspond to the noninteracting spectrum. 
}
\label{Fig:Akw}
\end{figure}

\section{Equilibrium}\label{Sec:Equilibrium}

In the following we will focus on  the parameter regime relevant for
La$_2$CuO$_4$, as obtained from LDA calculations and the constrained RPA downfolding to the low-energy space composed of the $d$ and $p$
orbitals, with parameters $U_{dd}=5.0$ eV, $U_{dp}=2.0$ eV, $t_{dp}=0.5$ eV, $t_{dd}=-0.1$ eV, $t_{pp}=0.15$ eV, and $\Delta_{pd}=-3.5$ eV,  see Refs.~\onlinecite{werner2015Dynam,hansmann2014} for \textit{ab-initio} estimates. In all calculations, we
set the inverse temperature to $\beta=5.0$ eV$^{-1}\approx 1/2000$ K$^{-1}$, which is above the
N\'{e}el temperature. We have employed this high temperature
due to the limitations of the NCA approximation, which prevent us from stabilizing calculations at lower temperatures. As the studied system has a 2 eV gap, the thermal fluctuations across the gap even at this elevated temperature can be neglected and we do not expect qualitative changes for lower temperatures until the N\'{e}el temperature is reached. We will compare results for two different approximations, namely, $GW$+EDMFT and
Hartree-Fock+EDMFT (HF+EDMFT). As only $GW$+EDMFT includes nonlocal dynamical charge
fluctuations, the comparison between the two methods will illustrate the
role of nonlocal screening. 

In Fig.~\ref{Fig:Spectral}, we present the
orbital-resolved local spectral function
$A_{A}(\omega)=-\frac{1}{\pi}\text{Im}[G_{\text{loc},AA}(\omega)],$
for $A=d,p_x,p_y$ obtained with the two schemes. 
The lower Hubbard band and the $p$ orbital lie in the same energy range and are strongly hybridized due to the direct hopping  $t_{dp}$. The spectral function below the Fermi level is split into three distinct peaks, see discussion in Ref.~\onlinecite{zhang1988}:
(a) a peak at $\omega=-3.5$ eV of predominantly $p$ orbital character corresponding to the nonbonding orbital combination, (b) the antibonding band corresponding to the Zhang-Rice singlet around $\omega=-1.5$ eV, and (c) the bonding band pushed to lower energy, with a center at
$\omega=-5$ eV. The Zhang-Rice band has been originally identified as an antibonding singlet state in a typical cuprate setup~\cite{zhang1988}.  While our numerical approach properly captures the orbital nature of this state, the two-particle spin fluctuations would have to be included to capture the singlet nature of the wave function. A detailed analysis using a cluster extension of DMFT showed that the feedback of the spin fluctuations onto the single-particle spectrum leads to a narrowing of the Zhang-Rice band due to the polaronic effect~\cite{yin2008}.

The presence of a large gap suppresses the screening in equilibrium, which explains the very similar $GW$+EDMFT and HF+EDMFT results. (We have explicitly checked that this is not anymore the case in doped systems.) The weak screening effect is also evident in a small reduction of the effective impurity interaction $\mathcal{U}(\omega=0)-U^{dd}\approx -0.05$ eV, so that the static approximation is justified.

The momentum- and orbital-resolved equilibrium spectral function 
$A_{k}(\omega)$,
shown in Fig.~\ref{Fig:Akw}, reveals that the band of predominantly $p$ character is split into two distinct bands around the $X$ point. This splitting is present already in the noninteracting spectrum (black lines) and originates primarily from the direct $p_x$-$p_y$ hopping, although also the hopping between the $d$ and $p$ orbitals leads to a slight splitting. The hybridization between the $d$  and $p$ band (see admixture of blue and red color in Fig.~\ref{Fig:Akw}) is strongest between the $X$ and $M$ point and between the bonding band and the $p$ bands.

\section{Photo-excitation}\label{Sec:Excitation}
\subsection{Orbital occupation}
Now we turn to the relaxation dynamics after photoexcitation. We excite the system with a short pulse
\begin{equation}
E(t) = E_0e^{-4.6(t-t_0)^2/t_0^2} \sin(\Omega(t-t_0))
\end{equation}
with frequency $\Omega$ and amplitude $E_0$,  polarized along the (11) direction. The width of the pulse $t_0 = 2\pi n/\Omega$ is chosen such that the envelope accommodates $n = 2$ cycles, unless otherwise stated.   

\begin{figure}[t]
\includegraphics[width=1.0\linewidth]{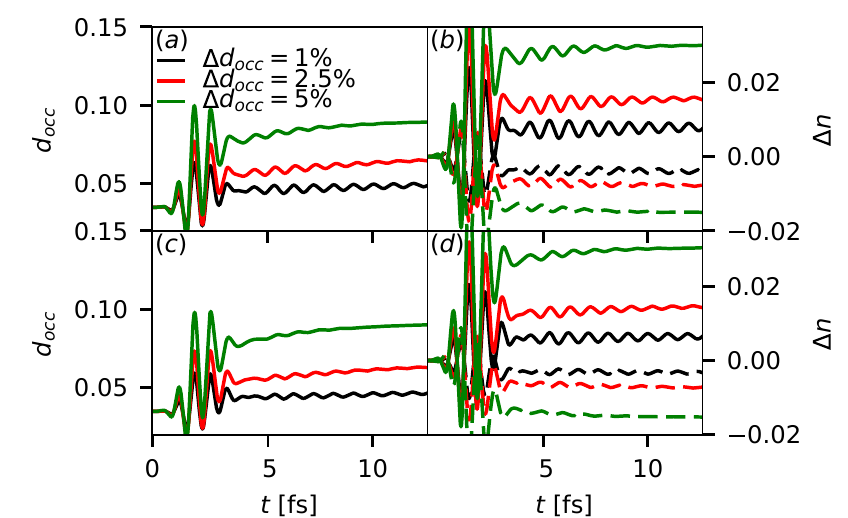}\\  
\caption{Time evolution of the change in the 
  double occupancy $d_\text{occ}$ (first columnn) and the orbital
  occupation of the $d$ (full line) and $p_x$ $(p_y)$ (dashed line) orbitals (second column). The data in the first (second) row have been obtained using the dipolar matrix element $|\text{D}|=0.3$ $e$a (0.1 $e$a), while the photon frequency is $\Omega=6.0$ eV.
  }
\label{Fig:docc}
\end{figure}

\begin{figure*}[t]
\includegraphics[width=0.75\linewidth]{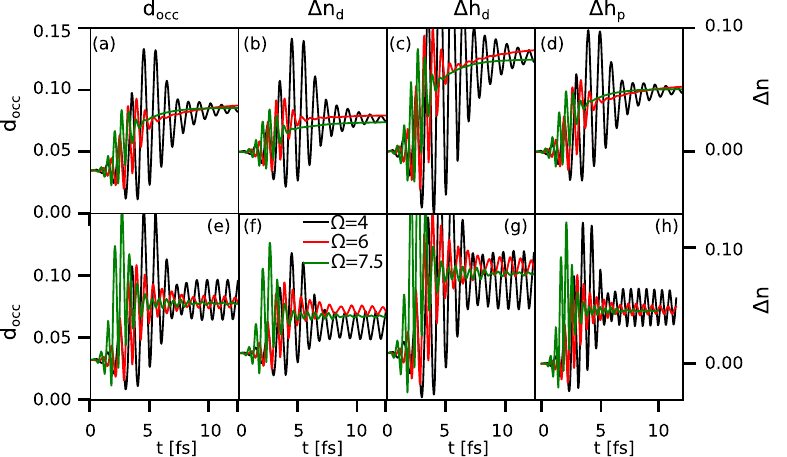}
\caption{Test of the orbital-selective excitation in the $GW$+EDMFT~(first row) and HF+EDMFT~(second row) description for transitions from the ZRS to the UHB ($\Omega=4.0$ eV), the $p$-band to the UHB ($\Omega=6.0$ eV) and the bonding band to the UHB ($\Omega=7.5$ eV). The first column presents the time evolution of the change in the 
  double occupancy $d_\text{occ}$. The second column shows the change in the orbital occupation of the $d$ orbital, $\Delta n_d$. The change in the $p$-orbital occupation is given by $\Delta n_p=-0.5\Delta n_d$. The last two columns show the  density of holes in the lower Hubbard band and the $p$ bands. 
   The dipolar matrix element is fixed to $|\text{D}|=0.3 $ $e$a.}
\label{Fig:docc_om}
\end{figure*}

The electric field pulse creates charge-transfer excitations and results in an increase of the double occupancy $d_\text{occ}=\langle n^d_\uparrow n^d_\downarrow \rangle$ of the $d$ orbitals, as evidenced in the time evolution illustrated in Figs.~\ref{Fig:docc}(a) and (c). In the following, we will adjust the strength of the pulse ($E_0$) to fix the photodoping at time $t=12$ fs to be $\Delta d_\text{occ}$= 1$\%$,~2.5$\%$, and 5$\%$.  The magnitude of the dipolar matrix element determines the ratio between direct interband excitations and Landau-Zener-like tunneling due to the intraband acceleration. The latter can be seen from the comparison of the  contributions from the Peierls term in Eq.~(\ref{Eq:current_Peierls}) and the dipolar term in Eq.~(\ref{Eq:current_dipol}) to the photoinduced current. Because the dipolar and the Peierls contribution have different interorbital matrix elements, it is expected that if one or the other of the two processes dominantes, the orbital population~(density of holes in the $d$ and $p$~bands) or the photo-induced current (optical conductivity) may be different after the photo excitation. 
We investigate the effect of the dipolar matrix element by comparing two values, namely $|\text{D}|=0.3$ $e$a and $|\text{D}|=0.1$ $e$a, at fixed photo-doping. In order to reach the same photo-doping concentration, a much larger field strength $E_0$ is needed in the case of the smaller dipolar matrix element. Surprisingly, the final occupation of the orbitals for fixed photo-doping, see Figs.~\ref{Fig:docc}(b) and~\ref{Fig:docc}(d), is almost independent of the dipolar matrix element. This illustrates that the state reached after the photodoping excitation is essentially determined by $d_\text{occ}$ and does not depend on the details of the excitation process or the value of the dipolar matrix element. 

To further confirm this finding, we investigate the effect of different pulse frequencies. In particular, we will focus on excitations from the ZRS to the UHB ($\Omega=4.0$), from the band with predominant $p$ character to the UHB ($\Omega=6.0$) and the bonding band to the UHB ($\Omega=7.5$).  A similar analysis has already been presented in Ref.~\onlinecite{golez2019}, with the conclusion that an orbital selective excitation of charge carriers is not possible. Here we make this test stronger by using pulses with a larger number of cycles, $n_p=4$, corresponding to a narrower energy distribution of the pump pulse. The results are presented in Fig.~\ref{Fig:docc_om}, where we also provide a direct comparison to the HF+EDMFT results. As the absorption is reduced in HF+EDMFT, a stronger pulse strength has to be applied to reach a fixed number of 
 doublons after the pulse.
 The excitation from the ZRS to the UHB produces pronounced oscillations in both approximations, see Figs.~\ref{Fig:docc_om}(a) and (b). For higher frequencies the inclusion of the nonlocal fluctuations strongly damps the long-lived oscillations in comparison to HF+EDMFT. Within $GW$+EDMFT the long-time orbital occupation of the $d$ orbital is almost independent of the pump pulse frequency, which further confirms the statements in Ref.~\onlinecite{golez2019}. In contrast, in HF+EDMFT, the $d$-orbital occupation exhibits a more pronounced pulse-frequency dependence.

The change in the $d$ occupancy, $\Delta n_d$, is related to the change in the $p$ occupancy, $\Delta n_p$, by $\Delta n_d=-2\Delta n_p$. The change in the double occupancy $\Delta d_\text{occ}$ corresponds to the total amount of charge transferred across the gap, and hence the change in the number of empty $d$-sites can be extracted as $\Delta h_d=\Delta d_\text{occ}+2\Delta n_p$. Since the $p$ orbitals are initially fully filled, the hole density on the $p$ orbitals (corresponding to singly occupied sites) can be estimated as $\Delta h_p=0.5 \Delta n_d$, see Fig.~\ref{Fig:docc_om} for the dynamics of both types of holes.   Similarly to the dynamics of the occupation, the dynamics of the density of holes in the lower Hubbard band shows a strongly oscillatory behavior in both approximations only for the lowest frequency $\Omega=4.0$. For high-frequency excitations, see Figs.~\ref{Fig:docc_om}(c) and~\ref{Fig:docc_om}(g), the density of $d$ holes in the lower Hubbard band is small and the long-time dynamics is essentially constant within HF+EDMFT. In contrast, the $GW$+EDMFT result demonstrates a clear relaxation toward an enhanced orbital occupancy of the $d$ orbital. Due to the strong oscillations in the number of $d$ and $p$ holes in HF+EDMFT, it is hard to discuss the orbital selectivity of holes. However, it is clear that the deviation between different excitations is much larger than in $GW$+EDMFT. Within the latter approximation, the system relaxes to a state with  practically identical occupations of the orbitals and similar hole concentrations for all excitation frequencies. The explicit relaxation dynamics of the holes can be monitored by the greater component,  which will be discussed in the following section. 

\subsection{Spectral function and occupation}\label{Sec:PES}
Now we turn to the dynamics of the spectrum and the population dynamics.
The time-dependent spectral function is defined as a partial Fourier transform over the difference in time
$A(\omega,t)=-\frac{1}{\pi}\text{Im}\int_{t}^{t+t_\text{cut}} dt' e^{\mathrm{i} \omega (t'-t) }G^{R}(t',t),$ with a cutoff $t_{\text{cut}}=8$ fs. This allows us to compute the spectrum with a time-independent resolution up to relatively long times. We have checked that the precise choice of the cutoff does not qualitatively affect the dynamics of the spectral features discussed below.

\begin{figure*}[t]
\includegraphics{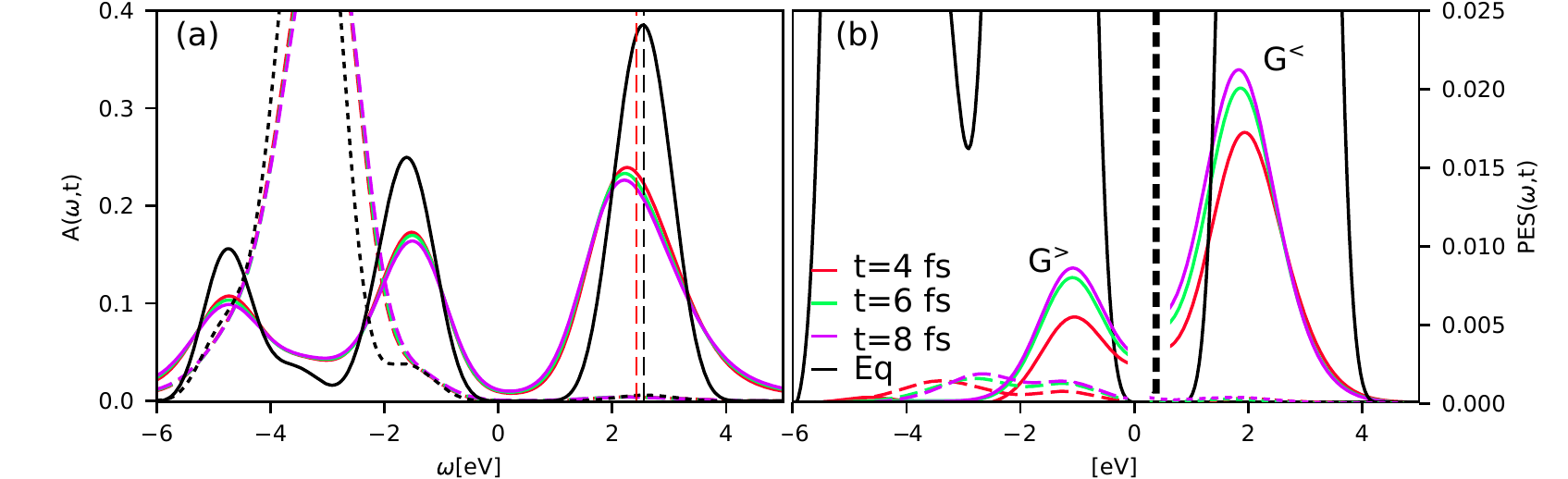} 
\caption{(a) Time evolution of the orbital-resolved spectral function $A(\omega,t)$ for equilibrium (black) and fixed delay times of 4, 6 and 8 fs.
The $p$ ($d$) orbitals correspond to the dashed (full) lines. (b) The analogous plot for the greater component at negative frequencies $G^>(\omega<0,t)$ and lesser component for positive frequencies $G^<(\omega>0,t)$, with the black line representing the equilibrium spectral function $A(\omega)$. Both panels show $GW$+EDMFT results.
The pump pulse amplitude is chosen such that approximately 5$\%$ photodoping is created at the frequency $\Omega=7.5$ eV, while the dipolar moment is fixed to $|\text{D}|=0.3$ $e$a.
}
\label{Fig:PES}
\end{figure*}

\begin{figure*}[t]
\includegraphics{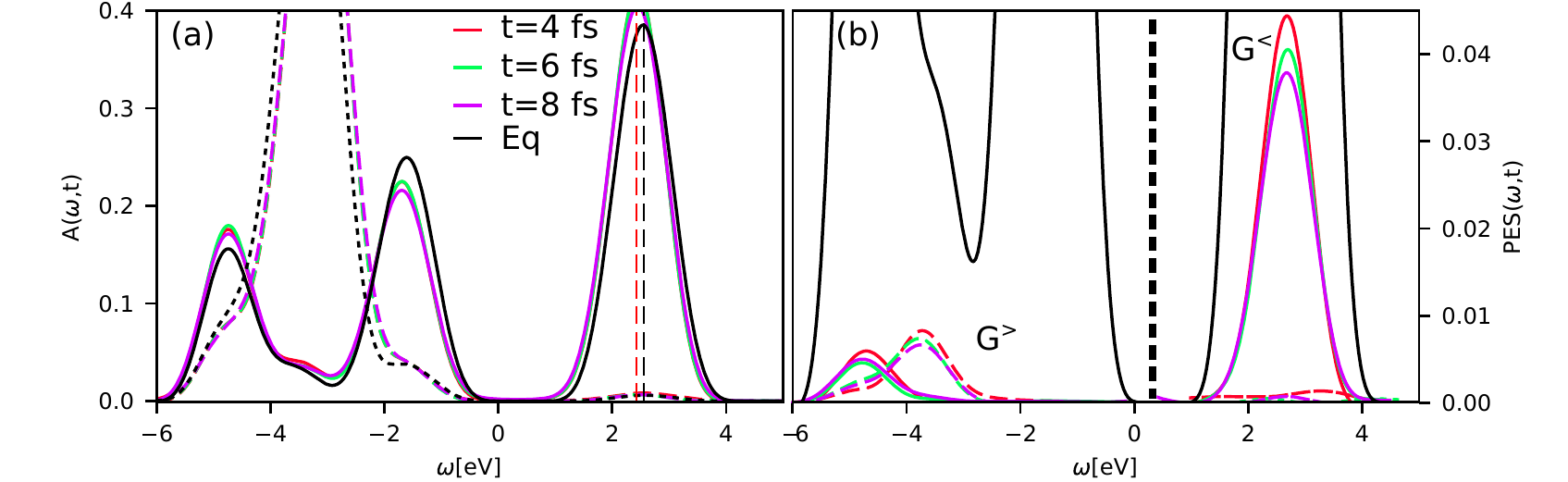}
\caption{Analogous plot to Fig.~\ref{Fig:PES} for the time-dependence in the HF+EDMFT approximation.
}
\label{Fig:PES_HF}
\end{figure*}

In the following, we will compare the photoinduced changes of the spectrum for $GW$+EDMFT~[Fig.~\ref{Fig:PES}(a)] and 
HF+EDMFT~[Fig.~\ref{Fig:PES_HF}(a)]. The band-gap renormalization is present in both cases, however it is much  larger in $GW$+EDMFT. Moreover, the $GW$+EDMFT result exhibits strong broadening of all spectral features. We can distinguish two processes leading to the band gap renormalization: (a) the static Hartree-shift due to the Coulomb interaction $U_{dp}$ between holes in the $p$ orbitals and doublons in the $d$ orbital, $\Delta \Sigma^{H}_{dd}=(U_{dd}-2U_{dp})\Delta n_d$~\cite{cilento2018,peli2017,lantz2017}, where the factor of 2 originates from the number of nearest-neighbor $p$ orbitals and we have used the conservation of the total charge $\Delta n_d =-(\Delta n_{p_x}+\Delta n_{p_y})$, (b) the photoinduced enhancement of screening, which reduces the effective interaction. To demonstrate that the main contribution to the band-gap renormalization originates from the dynamical screening, we mark the static Hartree shift $\Delta\Sigma_{H}$  by a vertical red line in Figs.~\ref{Fig:PES}(a) and~\ref{Fig:PES_HF}(a). Within HF+EDMFT, the Hartree shift perfectly describes the photoinduced changes, while for $GW$+EDMFT  it only accounts for a small part of the observed band renormalization, see also the discussion in Ref.~\onlinecite{golez2019}.

\begin{figure*}[t]
\includegraphics{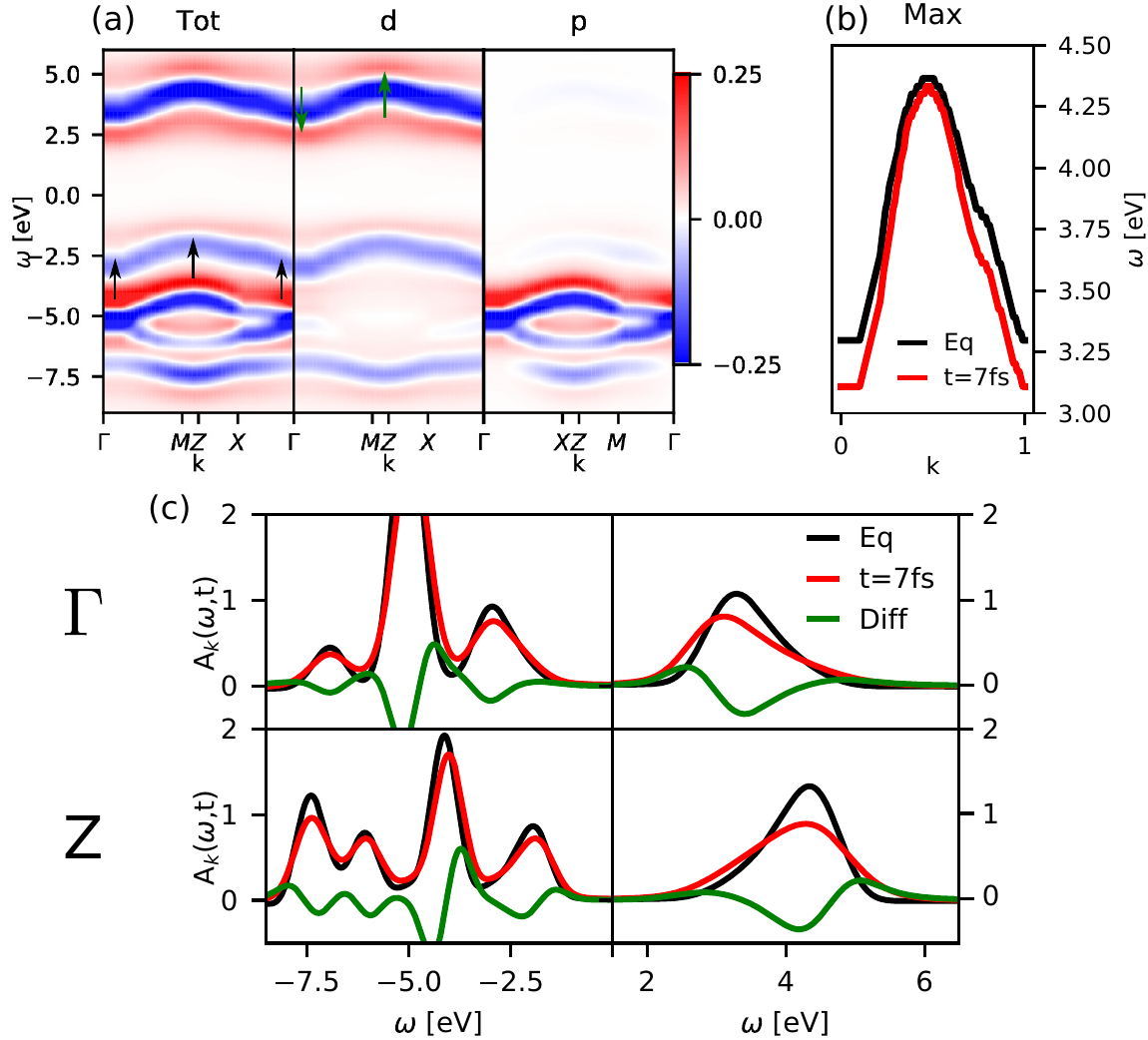}
\caption{(a)~Photo-induced difference in the momentum- and orbital-resolved spectral function $\Delta A_{k}(\omega,t)=A_{k}(\omega,t)-A_{k}^\text{eq}(\omega)$ traced  over all bands~(Tot) and projected onto the $d$ and $p$ orbitals. The black~(green) arrows in (a) indicate the corresponding momentum-dependent changes of the $p$-band~(UHB). (b) The maximum of the spectral function in the UHB before (black) and after photo-excitation (red). (c) Slices of the momentum-dependent spectrum at the $\Gamma$ and Z point in equilibrium~(black), after the pulse~(red), and their difference~(green). The first~(second) column shows a zoom of the spectrum below~(above) the chemical potential.  The pump pulse intensity is $E_0=0.31$, the frequency $\Omega=6$ eV, and the dipolar moment is fixed to $|\text{D}|=0.3$ $e$a, which corresponds to 5\% photo-doping. All panels show $GW$+EDMFT results.
}
\label{Fig:dAkw}
\end{figure*}

Within $GW$+EDMFT, the local spectrum is also strongly broadened, see Fig.~\ref{Fig:PES}(a). The broadening  can originate from two
distinct mechanisms: (a) a change in the lifetime due to increased scattering and (b) photoinduced changes in the band dispersion~(velocities). To distinguish these two effects, we  analyze the change in  the momentum- and orbital-resolved spectral function with respect to its equilibrium value $\Delta A_{k}(\omega,t)=A_{k}(\omega,t)-A_{k}^\text{eq}(\omega)$, see Fig.~\ref{Fig:dAkw}. The momentum-dependent information shows that the change in the region of the UHB depends on momentum: the  spectral weight close to the $\Gamma$ point is shifting towards the chemical potential, while spectral weight near the $Z$ point is shifting away from the chemical potential~[green arrows in the $d$-orbital panel of Fig.~\ref{Fig:dAkw}(a)]. A detailed comparison of the momentum cut at the $\Gamma$ point in Fig.~\ref{Fig:dAkw}(c) reveals asymmetric changes and a net shift of spectral weight towards the chemical potential at the lower edge of the UHB, see the green lines for the difference of the spectra in Fig.~\ref{Fig:dAkw}(c). The  opposite trend is seen at the Z point. The latter difference could originate either from the photoinduced changes in the Fock contribution, which affects the bandwidth of the Hubbard band\cite{ayral2017}, or the $GW$ contribution originating from the momentum-dependent scattering with the plasmonic excitations. We have checked that the change in the bandwidth due to the Fock term after the excitation is at least an order of magnitude smaller than the shift in the momentum-dependent spectrum. For the UHB, the latter is presented in Fig.~\ref{Fig:dAkw}(b) as a photoinduced change in the maximum of the spectrum. Besides the change in the dispersion, the total change in the spectrum is accompanied by a broadening, which is roughly momentum independent. This leads to the conclusion that the  photoinduced changes in the UHB and gap originate from the combination of both, namely a broadening of the spectrum, and changes in the dispersion.

In contrast, the Zhang-Rice singlet exhibits only a slight broadening, without any clear changes in the dispersion. Moreover, the main part of the  $p$ band  shows an almost perfectly rigid band shift with minor lifetime effects, in agreement with the change in the local spectrum, see black arrows in Fig.~\ref{Fig:dAkw}(a). Because the shift is rigid, the effect is also seen in the local~($k$-integrated) spectra, see Fig.~\ref{Fig:PES}(a). This is confirmed by the behavior  of the photoinduced difference at $\omega=-5.0$ in Fig.~\ref{Fig:dAkw}(c), which shows a symmetric oscillation-like shape. An interesting observation is that the bonding band shows a dichotomy between the $d$ and $p$ contribution, where the latter exhibits a momentum dependent broadening, while the broadening of the former is almost momentum independent, compare the $d$ and $p$ spectra in Fig.~\ref{Fig:dAkw}(a). The comparison of the differences in the corresponding 
spectral peaks 
at the $\Gamma$ and Z points, see Fig.~\ref{Fig:dAkw}(c), confirms that these quasi-particles are only broadened without a clear shift.

To investigate the population dynamics we compute the time- and orbital-resolved lesser component $G^<(\omega,t)$, which measures the occupation dynamics as a function of the probe time $t$ and the energy $\omega$
\begin{align}
\label{PES}
G^<(\omega,t)= \frac{1}{\pi}\text{Im} \int_{t}^{t+t_\text{cut}} dt' e^{\mathrm{i} \omega (t'-t) }G^{<}(t',t).
\end{align}
This expression is similar to the formula for the photoemission spectrum~\cite{freericks09}, but does not contain the pulse envelope functions. This simple forward integration allows us to analyze longer times.

We choose to excite the system with a pulse which is resonant to the transition between the bonding band and the UHB~($\Omega$ = 7.5 eV) to clearly resolve the hole dynamics. We have checked that the dynamics of holes is qualitatively similar for the resonant excitation between the predominantly $p$ band and the UHB. In Figs.~\ref{Fig:PES}(b) and~\ref{Fig:PES_HF}(b), we compare the evolution of holes in the $GW$+EDMFT and HF+EMDFT approximation. The lesser~(greater) component is plotted for $\omega>0$ $(\omega<0).$

 After the pump pulse the occupation dynamics shows a relaxation of the doublons to the lower edge of the UHB within $GW$+EDMFT, see $G^<$ curves in Fig.~\ref{Fig:PES}(b). In contrast, the doublon distribution does not relax within HF+EDMFT, see Fig.~\ref{Fig:PES_HF}(b). Interestingly, a small number of $p$ electrons is also present in the UHB due to the direct hybridization between the doublons on the $d$-orbital and the electrons on the $p$-orbitals. 

\begin{figure*}[t]
\includegraphics{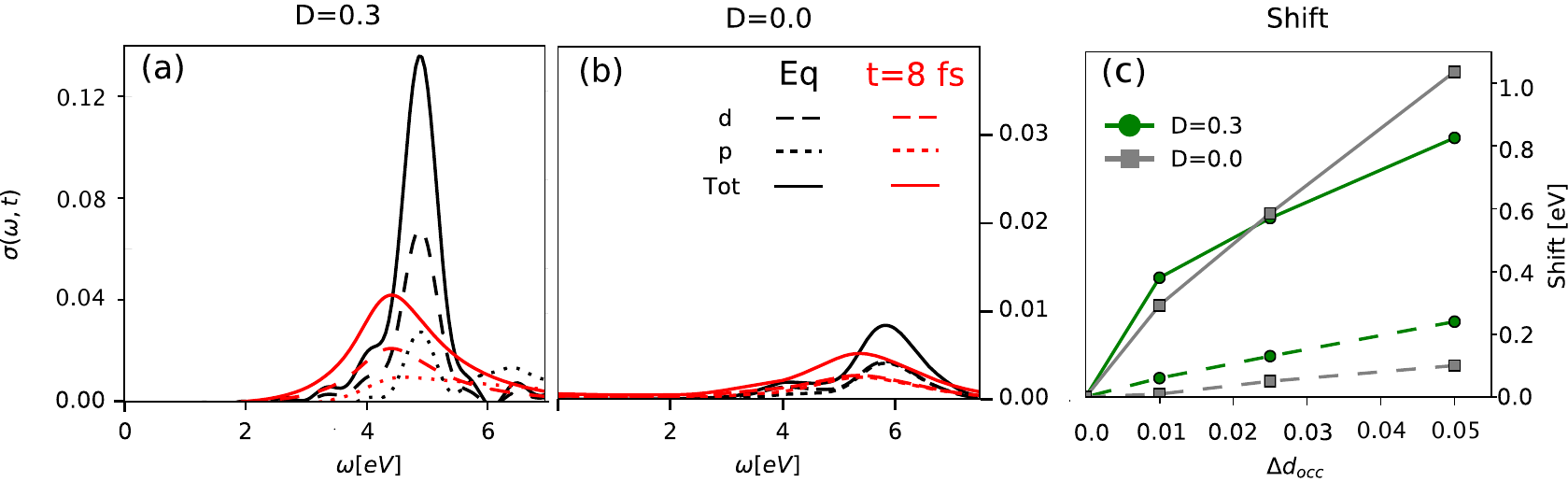}
\caption{
Orbital-resolved and total optical conductivity in equilibrium (black) and after the pump pulse (red) for the dipolar matrix elements (a)~$|\text{D}|=0.3$ $e$a and (b)~$|\text{D}|=0.0$ $e$a. (c) Gap size renormalization at the position of the half-maximum in the optical conductivity (full lines) as a function of the density of photoexcited doublons and comparison with the renormalization expected from the static Hartree shift $\Delta \Sigma^{H}_{dd}$~(dashed) for two values of the dipolar matrix element $|\text{D}|=0.0,0.3$ $e$a. The frequency of the excitation is $\Omega=6$ eV.
All panels show $GW$+EDMFT results.}
\label{Fig:optD}
\end{figure*}

In order to follow the dynamics of the holes, we will now analyze the greater component $G^>(\omega,t)=\frac{1}{\pi}\text{Im}\int_{t}^{t+t_\text{cut}} dt' e^{\mathrm{i} \omega (t'-t) }G^{>}(t',t).$ 
In both approximations, the photo-excitation creates a substantial density of holes on the $p$ orbitals. Within $GW$+EDMFT, see Fig.~\ref{Fig:PES}(b), these $p$ holes quickly relax to the upper band edge. The weight of the $d$ holes increases, which is the result of relaxation processes including inter-band scattering and impact ionization\cite{werner2014}. These two processes are responsible for a substantial increase in the density of $d$ holes in the LHB, compare Figs.~\ref{Fig:PES}(a) and~\ref{Fig:docc}. In contrast, the holes are almost trapped within HF+DMFT, see Fig.~\ref{Fig:PES_HF}(b), as the interband relaxation is highly suppressed without charge fluctuations. Assuming that the pulse creates holes at similar energies in both approximations, the comparison furthermore shows that due to the nonlocal and interband charge fluctuations a significant fraction of holes already relaxes during the pulse.

The experimental renormalization of the nonbonding oxygen $p$
band has been recently reported in a time-resolved ARPES study of optimally doped Y-Bi2212 single crystals, see Ref.~\onlinecite{cilento2018}. As we are comparing a doped system in the experiment with a half-filled system in theory, a comparison close to the chemical potential is not possible. Therefore, we will focus on the dynamics of the nonbonding oxygen $p$ band close to the anti-nodal point. In the experiment, the dynamics of the oxygen band is characterized by a long-lived spectral weight variation and the comparison with Fig.~\ref{Fig:PES}(a) for the $p$ band close to the X point reveals a qualitatively similar behavior. While experimentally, the renormalization of the nonbonding $p$ band is only reported close to the X point, our modeling predicts an almost momentum-independent renormalization. A more detailed comparison between theory and experiment would be desirable to understand the momentum-dependent renormalization in the whole BZ.

\subsection{Optical conductivity}

Optical spectroscopy has been extensively used to study the photoinduced dynamics in charge-transfer insulators.\cite{giannetti2016} The pioneering experiments on cuprates have shown a clear band gap renormalization, which at the time was attributed to photoinduced in-gap states.\cite{matsuda1994} Subsequent experiments confirmed the effect and pointed toward a multiband origin of the gap renormalization.\cite{okamoto2011,okamoto2010,novelli2014} Here we analyze the optical absorption by explicit simulations of the pump and probe pulse and extract the photoinduced current as
the difference of the current with and without a probe pulse,  
$j_\text{probe}=j_\text{pump+probe}-j_\text{pump}.$ For a weak probe pulse, we evaluate the optical
conductivity  as the ratio  
\begin{equation}
\sigma(\omega,t_p)=j_\text{probe}(\omega,t_p)/E(\omega,t_p),
\end{equation} 
where $X(\omega,t_p)=\int_0^{t_\text{cut}} ds X(t_p+s) e^{-\I \omega s - \eta s}$ is the Fourier transform of $X=j_\text{probe}$ or $E,$ and
$t_p$ is the start of the probe pulse. Due to the long-lived oscillations in the photoinduced current, a rather large broadening $\eta=0.2$ is used to avoid artifacts from the finite time-window Fourier transform. This procedure avoids the calculation of the current-current correlation function including vertex corrections.

In order to address the role of the dipolar matrix element $\text{D}$ on the dynamics of the optical conductivity we compare results for $|\text{D}|=0.0$ $e$a and $0.3$ $e$a. In the following, we apply both the pump and probe pulses in the (11) direction. Different dipolar matrix elements $\text{D}$ modify the equilibrium optical conductivity, compare Fig.~\ref{Fig:optD}(a) and~\ref{Fig:optD}(b). In both cases, the optical signal is composed of two characteristic features corresponding to either the transitions between the Zhang-Rice singlet and the UHB (smaller peak at $\omega\approx 4$ eV) or from the band with predominant $p$ character to the UHB. The position of the latter shows a substantial shift to lower energies as the dipolar matrix element increases. This suggest that the main contribution of the dipolar current comes from slightly lower energies than the Peierls contribution, which we have confirmed by analyzing both the intraband and interband contributions to the current~(not shown). Moreover, the height of the peaks is strongly increased for  $|\text{D}|=0.3$ $e$a compared to $|\text{D}|=0$ $e$a. 
This is easily understood as the bigger dipolar matrix element leads to a higher transition probability between bands with different orbital character.  We have checked that the heights of the peaks do not simply scale with the size of the dipolar term. The orbital-resolved optical conductivity shows that the photo-induced current through the $p$ and $d$ orbitals is almost the same. This is a consequence of a  small direct hopping between the $p_x$ and $p_y$ orbitals, namely, t$_{pp}=0.15$ eV. Hence, the current is mainly originating from the hopping between $p$ and $d$ orbitals. 
 
After the photoexcitation, the edges of these peaks are shifted to lower energies and furthermore, the peaks get broadened in analogy to the single-particle spectra. Both effects are clearly seen for the main peak for both values of the dipolar matrix element. While the height of the main peak is reduced in both cases, the effect is strongly enhanced for $|\text{D}|=0.3$ $e$a. This can be intuitively understood as a population effect since after the photoexcitation the presence of doublons in the UHB partially prevents a direct dipolar excitation.  The peak at $\omega=3.5$ eV, which corresponds to transitions between the Zhang-Rice singlet and the UHB, is strongly smeared out and can only be recognized as a shoulder-like structure. The latter could be an effect of a rather large broadening $\eta=0.2$ and therefore we cannot resolve small photoinduced features associated with transitions between the Zhang-Rice singlet and the UHB after the photoexcitation. The orbital-resolved optical conductivity shows that the changes in the spectrum are comparable for all orbitals.

We extracted the photoinduced shifts for different photodopings from the change in the half-width at half-maximum for fixed dipolar matrix elements, see Fig.~\ref{Fig:optD}(c). The latter is compared to the static shift determined  from the expression for the Hartree shift $\Delta \Sigma^{H}_{dd}=(U_{dd}-2U_{dp})\Delta n_d$, which only depends on the number of the photo-induced doublons. We conclude that the static shifts are generally much smaller and their increase slightly depends on the dipolar matrix element, see Fig.~\ref{Fig:optD}. In contrast, the total shift has a stronger dependence on the dipolar moment. Without dipolar moment~($D=0.0$), the bandgap shift with respect to the photo-doping density  exhibits almost a linear dependence, while for finite dipolar moment the dependence becomes strongly nonlinear. Since the photo-induced changes for the different dipolar matrix elements $|\text{D}|=0.0$ and $0.3$ $e$a show a qualitatively and quantitatively different dynamics [compare Figs.~\ref{Fig:optD}(a) and~\ref{Fig:optD}(b)], it will be crucial to determine the dipolar matrix elements from \textit{ab-initio} calculations for comparison with experiments.

\section{Conclusions}\label{Sec:Conclusions}
We have discussed and demonstrated the nonequilibrium generalization of the multiband $GW$+EDMFT formalism by considering the Emery model  relevant for La$_2$CuO$_4$. This approach is based on a multiscale modeling, where the low-energy correlated $d$ orbitals are treated within $GW$+EDMFT, while the higher-lying $p$ orbitals are described with $GW$~(in $GW$+EDMFT) or Hartree-Fock~(in HF+EDMFT). We provided a detailed description of the downfolding formalism and the mapping to the effective impurity problem, which we solved within the noncrossing approximation~(NCA). 

We discussed the coupling of the electrons with classical light in the case of multiband systems. Starting from the continuum description and the minimal coupling, and after performing the Power-Zienau-Wolley transformation\cite{loudon2000}, we projected the Hamiltonian to the Wannier orbitals. A Peierls-like transformation then led to the  gauge-invariant formulation of the intraband acceleration described by a complex hopping parameter and additional interband dipolar transitions.

In the detailed analysis of the photodoped system, we addressed the three questions raised in the introduction. (i) We showed that the frequency of the pulse can lead to long-lived oscillations if the photoexcitation connects the Zhang-Rice singlet and the upper Hubbard band. For higher-frequency excitations, the systems' properties depend only on the strength of the excitations and even if predominantly $p$-like charge carriers are excited, they quickly relax toward the 
band edge (Zhang-Rice singlet state). 
However, the latter process is only described within $GW$+EDMFT, while within HF+EDMFT the charge carriers are almost frozen within the band. This analysis implies that the dynamical charge fluctuations speed up not only the intraband relaxation but also the interband relaxation. (ii) The size of the dipolar matrix element determines the amount of energy absorbed from the high-frequency excitations on the fs timescale. However, if we fix the amount of photodoping by adjusting the pulse strength, the long-time relaxation dynamics and photoinduced changes are agnostic to the excitation protocol. (iii) The photoexcitation leads to modifications of the bands and the charge-transfer gap. In the local spectral function, the band-gap renormalization and lifetime effects can be clearly distinguished. Moreover, the momentum- and orbital-resolved spectral function reveals that the main origin of these features is a nontrivial interplay between a photoinduced change of the effective velocity, $k$-dependent broadening of the UHB and a rigid shift of the $p$ band. (iv) The charge-transfer gap renormalization is also evident in the optical conductivity. First, the size of the band shift depends on the dipolar matrix element. Moreover, the width of the main peak and the visibility of the Zhang-Rice like structure after the pulse
qualitatively depends on the size of the dipolar moment.

An alternative approach to the formalism presented in this work is a recent hybrid implementation of the time-dependent density-functional theory~(TDDFT+U), which has been applied to photoexcited NiO~\cite{tancogne2018}.  Similar band shifts and  effectively reduced interactions have been observed, but a direct comparison is difficult due to the lack of spectral information in TDDFT+U. As the authors in Ref.~\onlinecite{tancogne2018} showed an interesting spectrum of high-harmonics, such an analysis would provide a useful test for the proposed formalism. The photoinduced changes in the screening spectrum~\cite{golez2017,golez2015} can lead to an enhanced recombination of charge carriers and consequently a larger high-harmonic response~\cite{murakami2018}. 

Apart from these insights into the nonequilibrium properties of charge-transfer insulators, our work represents an important step in the development of {\it ab-initio} simulation approaches for strongly correlated systems in nonequilibrium states. The $GW$+EDMFT method implemented here features a fully consistent treatment of correlation and screening effects, and can  be combined with material-specific input obtained within a multitier approach analogous to the scheme recently demonstrated in equilibrium systems in Refs.~\onlinecite{werner2015Dynam,boehnke2016,nilsson2017}.
 
\acknowledgements
D.G. acknowledges insightful discussions with Lewin Boehnke, Yuta Murakami and Michael Sch\" uler.  The calculations have been performed on the Beo04 and Beo05 clusters at the University of Fribourg and the Rusty cluster at the Flatiron Institute. DG and PW acknowledge support from SNSF Grant No. 200021\_165539 and ERC Consolidator Grant No. 724103. ME acknowledges financial support from ERC Starting Grant No. 716648. The Flatiron Institute is a division of the Simons Foundation.

\bibliography{bibtex/tdmft,bibtex/Polarons,bibtex/Books,bibtex/lightmatter}

\end{document}